\documentclass[onecolumn,showpacs,preprintnumbers]{revtex4}
\usepackage{amssymb}
\usepackage{amsfonts}
\usepackage{amsmath}
\usepackage{graphicx}
\usepackage{dcolumn}
\usepackage{bm}
\usepackage{epsfig}

\begin{document}

\title{Synchronous Lagrangian variational principles in General Relativity}
\author{Claudio Cremaschini}
\affiliation{Institute of Physics, Faculty of Philosophy and Science, Silesian University
in Opava, Bezru\v{c}ovo n\'{a}m.13, CZ-74601 Opava, Czech Republic}
\author{Massimo Tessarotto}
\affiliation{Department of Mathematics and Geosciences, University of Trieste, Via
Valerio 12, 34127 Trieste, Italy}
\affiliation{Institute of Physics, Faculty of Philosophy and Science, Silesian University
in Opava, Bezru\v{c}ovo n\'{a}m.13, CZ-74601 Opava, Czech Republic}
\date{\today }

\begin{abstract}
The problem of formulating synchronous variational principles in
the context of General Relativity is discussed. Based on the
analogy with classical relativistic particle dynamics, the
existence of variational principles is pointed out in relativistic
classical field theory which are either asynchronous or
synchronous. The historical Einstein-Hilbert and Palatini
variational formulations are found to belong to the first
category. Nevertheless, it is shown that an alternative route
exists which permits one to cast these principles in terms of
equivalent synchronous Lagrangian variational formulations. The
advantage is twofold. First, synchronous approaches allow one to
overcome the lack of gauge symmetry of the asynchronous
principles. Second, the property of manifest covariance of the
theory is also restored at all levels, including the symbolic
Euler-Lagrange equations, with the variational Lagrangian density
being now identified with a $4-$scalar. As an application, a joint
synchronous variational principle holding both for the non-vacuum
Einstein and Maxwell equations is displayed, with the matter
source being described by means of a Vlasov kinetic treatment.
\end{abstract}

\pacs{02.30.Xx, 04.20.Cv, 04.20.Fy, 05.20.Dd}
\maketitle


\section{Introduction}

Variational approaches to the standard formulation of General Relativity
(SF-GR) have been a popular subject of research since 1915 following the
original mathematical formulation of this type first established
independently by Einstein and Hilbert \cite{ein1,LL,hilb}. Nevertheless, the
topic still represents an open fundamental theoretical challenge because of
a number of critical issues which, to date, remain unsolved. Here we refer,
in particular, to the following key mathematical requirements:\ 1) the
fulfillment of the property of manifest covariance with respect to the group
of local point transformations; 2)\ the property of gauge invariance of the
related continuum Lagrangian formulations; 3) the proper prescription of the
functional setting to be adopted in these variational approaches in order to
fulfill properties 1) and 2).

The fact that these properties are actually mandatory in this context stems
from elementary physical/mathematical arguments. Indeed, regarding the first
requirement, the manifest covariance property is a necessary condition
following from the general covariance principle (GCP) which requires the
identical fulfillment of the property of covariance with respect to
arbitrary local point transformations (i.e., smooth and invertible local
coordinate transformations associated with the $4-$position). This implies,
in turn, that it must always be possible to cast an arbitrary relativistic
continuum field theory satisfying GCP, such as SF-GR, in a form which
fulfills the property of manifest covariance at all levels. In other words,
this means representing all involved quantities exclusively by means of $4-$%
tensor continuum fields, starting from the variational action functional,
its variational Lagrangian density, and including as well the corresponding
Lagrangian generalized coordinates, generalized velocities and momenta.\ As
a consequence, the related equations should all be manifestly covariant too,
including in particular the symbolic Euler-Lagrange equations determined in
terms of the variational Lagrangian. A GR theory which fulfills such a
property will be referred to as manifestly covariant.

The second requirement, by no means less physically-relevant, requires that
a gauge-representation is given to the relevant variational principles
holding in SF-GR. In turn, this demands the precise definition of the notion
of gauge invariance to be adopted in this context. This should follow by
analogy with the corresponding well-known flat space-time gauge theories
available for continuum fields. Nevertheless, such a property is generally
not met in variational GR approaches to be found in the literature.

The third requirement concerns the precise specification of the functional
class of variations which is involved in the prescription of the continuum
variational functionals. In principle, this should be determined in such a
way to meet the previous requirements 1) and 2). More precisely, this
refers\ to the prescription of: 1) the boundary conditions for the continuum
fields; 2)\ the variations for the same continuum fields; 3)\ the
variational action functional. In previous literature this issue has not
always being treated unambiguously. In fact, implicitly, most of the
literature adopts the same functional setting originally proposed by
Einstein (see for example Refs.\cite{ein1,fey}), and recalled below.

Given the fundamental importance of variational principles for the axiomatic
foundations of classical relativistic field theory, in this paper we intend
to address these topics in order to propose their possible consistent
solution. As indicated below, this is realized via the introduction of a new
representation for the variational action functional and of the
corresponding Lagrangian density associated with the continuum classical
fields. The resulting variational principle for the Einstein equations is
referred to as synchronous Lagrangian variational principle. As a
consequence, it is found that by construction the same Lagrangian density
exhibits the correct (i.e., manifest) covariance and gauge invariance
properties. The modified setting is shown to be of general validity for
classical fields. In particular, it applies also to the non-vacuum Einstein
equations, when classical sources represented by the electromagnetic (EM)
field and matter distributions are taken into account. For this purpose a
Vlasov kinetic description of the field sources is introduced, which permits
one to set the corresponding fluid descriptions on a rigorous basis (see for
example the related discussion in Refs.\cite{Cr2010,Cr2011,Cr2011a}).\textbf{%
\ }This feature is remarkable in itself. In fact, as shown below, it affords
the establishment of a joint synchronous variational principle in which all
classical sources for the Einstein and Maxwell equations, including the same
Vlasov equation, are dealt with by means of a single variational functional.

In the framework of SF-GR, the starting point is provided by the Einstein
equations for the physical observable associated with the structure of the
space-time, namely its symmetric 2-rank metric tensor, which arises in the
presence of external sources. Its solution coincides with the extremal field
indicated below, and therefore is denoted as $\overline{g}_{\mu \nu }=%
\overline{g}_{\mu \nu }\left( r\right) $. For the same reason, in the paper
all barred continuum fields will be intended either as functions of $%
\overline{g}_{\mu \nu }$ or independent extremal functions. Notice also that
here and in the following, $r$\ stands for the local functional dependence
in terms of the $4-$position $r^{\mu }$. Neglecting for the moment the
contribution associated with the cosmological constant $\Lambda $, such an
equation is given by the manifestly covariant 2nd-order PDE%
\begin{equation}
\overline{R}_{\mu \nu }-\frac{1}{2}\overline{R}\overline{g}_{\mu \nu }=\frac{%
8\pi G}{c^{4}}\overline{T}_{\mu \nu },  \label{eis}
\end{equation}%
to be supplemented by suitable boundary conditions \cite{gravi,wald}. This
problem determines uniquely the metric tensor, which can be equivalently
expressed either in terms of its covariant ($\overline{g}_{\mu \nu }$) or
its contravariant ($\overline{g}^{\mu \nu }$) components. The latter ones
are by construction the inverse of each other, so that they are related by
the condition $\overline{g}_{\mu \alpha }\overline{g}^{\mu \beta }=\delta
_{\alpha }^{\beta }$. The notation is standard. Thus, first $G$ is the
universal gravitational constant and $c$ is the speed of light in vacuum,
while $\overline{G}_{\mu \nu }\equiv \overline{R}_{\mu \nu }-\frac{1}{2}%
\overline{R}\overline{g}_{\mu \nu }$ denotes the symmetric Einstein tensor.
Furthermore, $\overline{R}\equiv \overline{g}_{\mu \nu }\overline{R}^{\mu
\nu }$ is the Ricci $4-$scalar, while $\overline{R}_{ik}$ is the related
Ricci curvature $4-$tensor%
\begin{equation}
\overline{R}_{ik}=\frac{\partial \overline{\Gamma }_{ik}^{l}}{\partial x^{l}}%
-\frac{\partial \overline{\Gamma }_{il}^{l}}{\partial x^{k}}+\overline{%
\Gamma }_{ik}^{l}\overline{\Gamma }_{lm}^{m}-\overline{\Gamma }_{il}^{m}%
\overline{\Gamma }_{km}^{l}.  \label{rirr}
\end{equation}%
Here $\overline{\Gamma }_{ik}^{l}$ denote a suitable coordinate
representation of the Levi-Civita connection functions, and hence are
inherently non-tensorial in character. More precisely, $\overline{\Gamma }%
_{ik}^{l}$\ identify the extremal Christoffel symbols, which are evaluated
with respect to the extremal field $\overline{g}_{\mu \nu }$ and\ defined as
usual as%
\begin{equation}
\overline{\Gamma }_{ik}^{l}=\frac{1}{2}\overline{g}^{lm}\left( \frac{%
\partial \overline{g}_{im}}{\partial x^{k}}+\frac{\partial \overline{g}_{mk}%
}{\partial x^{i}}-\frac{\partial \overline{g}_{ik}}{\partial x^{m}}\right) .
\label{ks}
\end{equation}%
Eq.(\ref{eis}) is completed by the metric compatibility condition in terms
of the covariant derivative of $\overline{g}^{\mu \nu }$:%
\begin{equation}
\overline{\nabla }_{\alpha }\overline{g}^{\mu \nu }=0,  \label{metric-com}
\end{equation}%
where by construction the covariant derivative operator $\overline{\nabla }%
_{\alpha }$ is defined with respect to the connections provided by the
extremal Christoffel symbols \cite{LL,wald}. Finally the symmetric
stress-energy tensor $\overline{T}_{\mu \nu }$ takes into account the
contribution of external sources. Its precise form is determined by the
prescription of the external mass distribution and the EM field.

\section{Goals}

The target of the paper is the construction of a new variational principle
for the non-vacuum Einstein equations, referred to here\textbf{\ }as \emph{%
synchronous Lagrangian variational principle}, which allows one to meet the
physical requirements indicated above. For this purpose, we find it useful
to present a brief introduction dealing with single-particle relativistic
dynamics,\ in which case the adoption of synchronous variational principles
is well-known. In fact, the distinction between synchronous and asynchronous
variational principles occurring in such a context permits one to identify
the route to follow also in the case of continuum field theory.\textbf{\ }%
Indeed, the physical analogy between the two problems provides a strong
argument in support of the validity of synchronous Lagrangian variational
principles also\ in the context of GR for relativistic field theory. For
greater generality, in the present work matter and charge current source
terms will be\ treated by means of a kinetic Vlasov description. A related
issue is therefore that of looking for a variational principle providing at
the same time also a synchronous variational principle of similar type for
the Vlasov equation. In particular, for the Vlasov equation the case of a
conservative mean-field vector field acting on single-particle dynamics will
be considered. The result allows one to determine a joint variational
principle for the non-vacuum Einstein and Maxwell equations, together with
the species Vlasov equations.

The structure of the presentation is as follows. In Section III and IV the
main physical motivations for the present investigation are discussed, which
arise from the analysis of standard literature approaches to the variational
formulation of GR. In Section V synchronous and asynchronous Lagrangian
variational principles of relativistic classical dynamics, together with
their basic features are recalled. Section VI deals with the formulation of
synchronous Lagrangian variational principles for the vacuum Einstein
equations (THM.2), while the extension to the non-vacuum case (i.e., in the
presence of sources) is considered in Section VII (THM.3). As an application
of the theory, in Section VIII first a joint synchronous variational
principle for the non-vacuum Einstein-Maxwell equations is established, with
the Vlasov source being included (THM.4). Then, a synchronous variational
principle is constructed for the Vlasov kinetic equation. Concluding remarks
are drawn in Section IX. Finally, in the Appendix a constrained asynchronous
variational principle is established (THM.1), which summarizes the
Einstein-Hilbert and Palatini variational approaches.

\section{Physical motivations}

As originally pointed out by Einstein and Hilbert \cite{ein1,LL,hilb}, the
non-vacuum equation (\ref{eis}) is intrinsically variational in character so
that, depending on the functional setting, it could be cast in terms of a
multiplicity of equivalent variational forms. In particular, it follows that
the same equation can always be represented by means of a suitable
Lagrangian variational principle (see below). In the Einstein and Hilbert
original approach as well as in the subsequent literature its construction
actually implicitly relies on the validity of a number of common underlying
hypotheses. These will be referred to in a short way as Einstein-Hilbert
(EH) axioms, being realized as follows:

\textit{EH axiom 1} - There exists a suitable functional class $\left\{
Z\right\} \equiv \left\{ Z=\left( Z_{1}...Z_{n}\right) ;f\left( Z\right)
=0\right\} $ of smooth real fields $Z_{i}\left( r\right) $, for $i=1,n$, to
be denoted as continuum Lagrangian generalized coordinates,\textbf{\ }which
are defined in the physical space-time $\mathbb{D}^{4}\equiv \left( \mathbf{Q%
}^{4},g_{\mu \nu }\right) $, where $\mathbf{Q}^{4}$ is a $C^{k}-$%
differentiable Lorentzian manifold, with $k\geq 3$, endowed with the metric
tensor $g_{\mu \nu }\left( r\right) $ yet to be determined. Furthermore, $%
Z_{i}\left( r\right) $ are required to satisfy appropriate boundary
conditions on a prescribed boundary $\partial \mathbb{D}^{4}$, namely $%
\left. Z_{i}\right\vert _{\partial \mathbb{D}^{4}}=Z_{i\mathbb{D}}$, where
the fields $Z_{i\mathbb{D}}$ are considered prescribed, namely for all $%
i=1,n $, to be the same for all fields $Z_{i}\left( r\right) $. Finally, the
same fields are also possibly subject to further suitable constraints,
symbolically represented here by equations of the form $f\left( Z\right) =0$%
. An arbitrary Lagrangian coordinate $Z_{i}\left( r\right) \in \left\{
Z\right\} $ will also be referred to as variational field.

\textit{EH axiom 2} - There exists a $4-$scalar functional $S\left( Z\right)
$ defined in $\left\{ Z\right\} $ and taking values in $%
\mathbb{R}
$ of the form%
\begin{equation}
S\left( Z\right) =\int_{\widehat{\mathbb{D}}^{4}}d^{4}x\sqrt{-g}L\left( Z,%
\mathcal{D}Z\right) .  \label{s}
\end{equation}%
This will be denoted as continuum Lagrangian action. Here the notation is as
follows. First, $\widehat{\mathbb{D}}^{4}\subseteq $ $\mathbb{D}^{4}$ is the
open interior subset of $\mathbb{D}^{4}$. Second, $d\Omega \equiv d^{4}x%
\sqrt{-g}$ identifies the $4-$scalar $4-$volume element, with
$d^{4}x\equiv \prod\limits_{i=0,3}dx^{i}$ being the canonical
measure and $g$ the determinant of $g_{\mu \nu }$, both to be
considered as variational
quantities, namely functions of the variational fields. Third, $L\left( Z,%
\mathcal{D}Z\right) $ is referred to as \textit{variational field Lagrangian}
and is identified with a real $4-$scalar smooth function of the variational
fields $Z$ and their \textquotedblleft generalized
velocities\textquotedblright\ $\mathcal{D}Z$ determined via an appropriate
differential operator $\mathcal{D}$ to be identified below.

\textit{EH} \textit{axiom 3} - The notion of variation $\delta Z_{i}$ must
be introduced for all the continuum Lagrangian fields $Z_{i}(r)$ belonging
to $\left\{ Z\right\} $. Thus, if $Z_{i}$ and $Z_{i}^{\prime }$ are two
arbitrary realizations of the $i-$th field, both belonging\ to\textbf{\ }$%
\left\{ Z\right\} $, then for all $Z_{i}$ the variation $\delta Z_{i}\left(
r\right) $ is prescribed by means of the scalar variation operator\textbf{\ }%
$\delta $\ so that%
\begin{equation}
\delta Z_{i}\left( r\right) =Z_{i}\left( r\right) -Z_{i}^{\prime }\left(
r\right) ,  \label{orig}
\end{equation}%
where $Z_{i}\left( r\right) -Z_{i}^{\prime }\left( r\right) $ is considered
infinitesimal and such that on the boundary it satisfies for all $i=1,n$ the
constraint equation $\delta Z_{i}\left( r\right) |_{\partial \mathbb{D}%
^{4}}=0$. This shall be referred to as variation of the function $%
Z_{i}\left( r\right) $. It follows that a generic variational field $%
Z_{1i}\left( r\right) \in \left\{ Z\right\} $ can always be expressed in
terms of the parametric representation%
\begin{equation}
Z_{1i}\left( r\right) =Z_{i}\left( r\right) -\alpha \delta Z_{i}\left(
r\right) ,
\end{equation}%
being $\alpha \in \left] -1,1\right[ $ an arbitrary real number which is
left invariant by the $\delta -$operator. Similarly, the variation of the
generalized velocities\textbf{\ }$\delta \mathcal{D}Z_{i}$ must also
generally be taken into account. This is identified with%
\begin{equation}
\delta \mathcal{D}Z_{i}\left( r\right) =\mathcal{D}Z_{i}\left( r\right) -%
\mathcal{D}Z_{i}^{\prime }\left( r\right) .  \label{first-1}
\end{equation}%
Requiring that the differential operators $\mathcal{D}$\ and $\delta $
commute, this implies that%
\begin{equation}
\delta \mathcal{D}Z_{i}\left( r\right) =\mathcal{D}\delta Z_{i}\left(
r\right) .  \label{first-2}
\end{equation}

\textit{EH axiom 4} - The continuum Lagrangian action functional $S\left(
Z\right) $ is assumed to satisfy the variational principle%
\begin{equation}
\delta S\left( Z\right) =0,  \label{action principle}
\end{equation}%
to hold for arbitrary variations $\delta Z\left( r\right) \equiv \left\{
\delta Z_{1},...\delta Z_{n}\right\} $. Here $\delta Z\left( r\right) $\
denotes the variation of the functional $S\left( Z\right) $,\ to be
identified with its Frechet derivative (see below). Then, provided the
differential operator $\mathcal{D}$\ is suitably defined, the related notion
of functional derivative $\frac{\delta S}{\delta Z_{i}}$ follows, for\ all $%
i=1,n$. Hence, the Euler-Lagrange equations corresponding to Eq.(\ref{action
principle}) are given, for all $i=1,n$, by%
\begin{equation}
\frac{\delta S}{\delta Z_{i}}=0,  \label{Euler-Lagrange equations}
\end{equation}%
where the variations are performed with respect to the \textit{variational
Lagrangian density, }namely%
\begin{equation}
\mathcal{L}\left( Z,\mathcal{D}Z\right) \equiv \sqrt{-g}L\left( Z,\mathcal{D}%
Z\right) .  \label{L-LL}
\end{equation}%
The latter is not a $4-$scalar. Nevertheless, by construction it is a
smoothly-differentiable ordinary real function, which depends on the
variables $\left( Z,\mathcal{D}Z\right) $. Notice, in fact, that the
variation of the action functional $\delta S\left( Z\right) $\ acts here, by
assumption, also on the coefficient $\sqrt{-g}$\textbf{\ }of the $4-$scalar
volume element which appears in the functional $S\left( Z\right) $ defined
according to Eq.(\ref{s}). For this reason Eq.(\ref{action principle}) will
be referred to here as an \emph{asynchronous variational principle.}

We stress that in the previous axioms the actual prescription of the
functional class $\left\{ Z\right\} $ and of the variational field
Lagrangian $L\left( Z,\mathcal{D}Z\right) $ remains arbitrary. We summarize
in the next paragraph the choices adopted for them in previous literature
for the Einstein equations, starting from the Einstein-Hilbert original
approach.

\subsection{Previous variational approaches}

In this section we briefly outline\ two examples of literature treatments
which actually permit the identification of a variational Lagrangian
density, namely the Einstein-Hilbert and the Palatini variational\textbf{\ }%
approaches. In both cases the variational field Lagrangian is identified with%
\begin{equation}
L=L_{EH}+L_{F},  \label{somma-lagrangiane}
\end{equation}%
where $L_{EH}$ denotes the Einstein-Hilbert vacuum field Lagrangian%
\begin{equation}
L_{EH}\left( Z,\mathcal{D}Z\right) \equiv -\frac{c^{3}}{16\pi G}R,
\label{LEH}
\end{equation}%
with $R$ being the Ricci $4-$scalar, which is assumed to be a function of
the variational fields defined in the functional class, while $L_{F}$\ is a
prescribed external source field Lagrangian. In the case the latter is
identified with the EM field, $L_{F}$ is denoted as $L_{EM}$ and considered
as a local smooth scalar function of the EM $4-$potential $A_{\mu }$. The
two approaches are characterized by different choices of the functional
class $\left\{ Z\right\} $. For definiteness, let us consider the case of
the vacuum Einstein equations. In the original Einstein-Hilbert variational
approach, also found in Ref.\cite{LL}, $\left\{ Z\right\} \equiv \left\{
Z\right\} _{E}$ is identified with the ensemble of generalized coordinates $%
g_{\mu \nu }(r)$ which are symmetric $4-$tensors in the indices $\mu ,\nu $,
and defined as%
\begin{equation}
\left\{ Z\right\} _{E}=\left\{
\begin{array}{c}
Z_{1}\equiv g_{\mu \nu }:g_{\mu \nu }\left( r\right) =g_{\nu \mu }\left(
r\right) \in C^{k}\left( \mathbb{D}^{4}\right)  \\
f_{1}\left( Z_{1}\right) =g^{\alpha k}g_{\beta k}-\delta _{\beta }^{\alpha
}=0 \\
\Gamma _{\alpha \beta }^{\mu }=\Gamma _{\left( C\right) \alpha \beta }^{\mu
}\left( g\right)  \\
g_{\mu \nu }\left( r\right) |_{\partial \mathbb{D}^{4}}=g_{\mu \nu \mathbb{D}%
}\left( r\right)  \\
\left. w^{\mu }\left( Z_{1},\mathcal{D}Z_{1}\right) \right\vert _{\partial
\mathbb{D}^{4}}=0 \\
\delta \left( d\Omega \right) =d^{4}x\delta \left( \sqrt{-g}\right) \neq 0%
\end{array}%
\right\} .
\end{equation}%
Notice that hereon $k\geq 3$ (to warrant the existence of $C^{1}$ solutions
for $g_{\mu \nu }$ in $\widehat{\mathbb{D}}^{4}$ which are continuous on $%
\partial \mathbb{D}^{4}$), $\mathcal{D}=\partial _{\mu }$ and $w^{\mu }$ is
the $4-$vector $w^{\mu }=g^{\alpha \beta }\delta \Gamma _{\left( C\right)
\alpha \beta }^{\mu }-g^{\alpha \mu }\delta \Gamma _{\left( C\right) \alpha
\beta }^{\beta }$, which depends both on $g_{\mu \nu }$ and its partial
derivatives.\textbf{\ }Here the constraint $f_{1}\left( Z_{1}\right) =0$\
warrants that the variational coordinates\textbf{\ }$g^{\mu \nu }(r)$\textbf{%
\ }and $g_{\mu \nu }(r)$\ raise and lower indexes (in particular in the
Lagrangian density $\mathcal{L}\left( Z,\mathcal{D}Z\right) $). Finally, $%
\Gamma _{\left( C\right) \alpha \beta }^{\mu }\left( g\right) $ are the
Christoffel symbols evaluated in terms of the variational field $g_{\mu \nu }
$, namely%
\begin{equation}
\Gamma _{\left( C\right) ik}^{l}\left( g\right) \equiv \frac{1}{2}%
g^{lm}\left( \frac{\partial g_{im}}{\partial x^{k}}+\frac{\partial g_{mk}}{%
\partial x^{i}}-\frac{\partial g_{ik}}{\partial x^{m}}\right) .
\label{christoffel-C}
\end{equation}%
These will be referred to as \textit{prescribed Christoffel symbols}. We
remark here that the choice of the boundary condition for $w^{\mu }$
involves the prescription of the partial derivative of $g_{\mu \nu }$ on the
boundary. An alternative possible definition of $\left\{ Z\right\} _{E}$
which avoids such a type of boundary condition can be found in Ref.\cite%
{wald}. In this case however the variational functional $S\left( Z\right) $
needs to be modified by means of the introduction of a surface contribution.
This feature actually prevents the introduction of a Lagrangian density as
indicated above.

The set of Euler-Lagrange equations associated with the Einstein-Hilbert
variational principle and based on the choices indicated above is
well-known. Written in the so-called \emph{symbolic representation}, namely
expressed in terms of the variational Lagrangian density, these are given by
the PDE%
\begin{equation}
\frac{1}{\sqrt{-g}}\frac{\partial \mathcal{L}}{\partial g^{\mu \nu }}=0,
\label{EL-1}
\end{equation}%
where $\mathcal{L}$ is now $\mathcal{L=}\sqrt{-g}L_{EH}\left( Z,\mathcal{D}%
Z\right) $ and the partial derivative with respect to the continuum
Lagrangian coordinate $g^{\mu \nu }$ must be performed keeping constant the
connections. We notice that $\sqrt{-g}$\textbf{\ }and $\mathcal{L}$\ are
separately not $4-$tensors, so that both\textbf{\ }$\frac{\partial \mathcal{L%
}}{\partial g^{\mu \nu }}$\textbf{\ }and $\frac{1}{\sqrt{-g}}\frac{\partial
\mathcal{L}}{\partial g^{\mu \nu }}$\textbf{\ }are not $4-$tensors too.
Hence, the\ symbolic representation of the Euler-Lagrange equation given by
Eq.(\ref{EL-1}) is not manifestly covariant. Nevertheless, once the explicit
calculation is performed, the previous equation delivers%
\begin{equation}
\frac{\partial L}{\partial g^{\mu \nu }}-\frac{1}{2}Lg_{\mu \nu }=0,
\end{equation}%
which coincides with the Einstein vacuum equation, and hence recovers the
property of manifest covariance.

The second approach to be mentioned is the one referred to in the literature
as the \textit{Palatini variational principle} \cite{gravi,wald}. This is
realized by considering both the metric tensor $g_{\mu \nu }$ and the
connections $\Gamma _{\alpha \nu }^{\mu }$ as independent continuum
Lagrangian coordinates. Lagrangian coordinates of this type will be referred
to as superabundant ones. In contrast, the ten independent components of the
variational metric tensor $g_{\mu \nu }(r)$ will be denoted as essential
Lagrangian coordinates. As a consequence, the functional class can now be
identified with $\left\{ Z\right\} \equiv \left\{ Z\right\} _{Pal}$,
represented by the ensemble of variational coordinates $g_{\mu \nu }(r)$ and
$\Gamma _{\alpha \nu }^{\mu }\left( r\right) $ which are both symmetric in
the lower indices $\mu ,\nu $, and is defined as%
\begin{equation}
\left\{ Z\right\} _{Pal}\equiv \left\{
\begin{array}{c}
\left[ Z_{1},Z_{2}\right] \equiv \left[ g_{\mu \nu }\left( r\right) ,\Gamma
_{\alpha \nu }^{\mu }\left( r\right) \right] \in C^{k}\left( \mathbb{D}%
^{4}\right) \\
f_{1}\left( Z_{1}\right) =g^{\alpha k}g_{\beta k}-\delta _{\beta }^{\alpha
}=0 \\
g_{\mu \nu }\left( r\right) |_{\partial \mathbb{D}^{4}}=g_{\mu \nu \mathbb{D}%
}\left( r\right) \\
\Gamma _{\alpha \nu }^{\mu }\left( r\right) |_{\partial \mathbb{D}%
^{4}}=\Gamma _{\alpha \nu \mathbb{D}}^{\mu }\left( r\right) \\
\delta \left( d\Omega \right) =d^{4}x\delta \left( \sqrt{-g}\right) \neq 0%
\end{array}%
\right\} ,  \label{Palatini}
\end{equation}%
with $k\geq 3$. The Euler-Lagrange equation corresponding to the functional
setting $\left\{ Z\right\} _{Pal}$\ and the same choice of the variational
Lagrangian given above follow in a similar way. In particular the variation
with respect to $\delta g^{\mu \nu \text{ }}$recovers again the symbolic
Euler-Lagrange equation given by Eq.(\ref{EL-1}). Instead, the remaining
extremal equation is obtained by considering the variation with respect to $%
\delta \Gamma _{\alpha \gamma }^{\beta }$. In symbolic form this is
expressed by the independent PDE%
\begin{equation}
\nabla _{\alpha }\left[ \frac{1}{\sqrt{-g}}\frac{\partial \mathcal{L}}{%
\partial R_{\mu \nu }}\right] =0.  \label{non-cov-metric}
\end{equation}%
Therefore,\ once again the symbolic Euler-Lagrange equations (\ref{EL-1})
and (\ref{non-cov-metric}) violate the property of manifest covariance.
Nevertheless, once the calculation of the partial derivatives is explicitly
carried out, the resulting PDEs recover the required covariance property. In
particular,\ as a result,\textbf{\ }it is immediate to show that Eq.(\ref%
{non-cov-metric}) simply reduces to Eq.(\ref{metric-com}).

In conclusion, it follows that both\ the Einstein-Hilbert and Palatini
variational approaches do not fulfill the property of manifest covariance of
the theory.\textbf{\ }Nevertheless, this feature is not-withstanding, as
both theories uniquely prescribe the vacuum Einstein equations in the
framework of mathematically-consistent approaches, i.e., being obtained in
terms of well-defined variational principles.

\subsection{An alternative constrained variational principle}

In this section we point out an alternative variational formulation for the
Einstein vacuum equations, which is still based on the adoption of an
asynchronous variational principle (see below) and is useful to display the
relationship with the Einstein and Palatini approaches recalled above.%
\textbf{\ }This is obtained by suitably prescribing the functional class $%
\left\{ Z\right\} $, while leaving unchanged the Einstein-Hilbert action. In
fact, one notices that formally Eq.(\ref{EL-1}) recovers identically the
Einstein vacuum equation provided the Ricci tensor $R_{\mu \nu }$ is kept
constant during variations, namely is identified with its extremal value $%
\overline{R}_{\mu \nu }$:%
\begin{equation}
R_{\mu \nu }=\overline{R}_{\mu \nu }.  \label{CONS}
\end{equation}%
Such a constraint can be satisfied adopting the method of the Lagrange
multipliers and introducing the functional class:%
\begin{equation}
\left\{ Z\right\} _{c}\equiv \left\{
\begin{array}{c}
\left\{ Z_{1}\left( r\right) ,Z_{2}\left( r\right) ,Z_{3}\left( r\right)
\right\} \equiv \left\{ g_{\mu \nu },R_{\mu \nu },\lambda ^{\mu \nu
}\right\}  \\
\overline{Z}_{2}\left( r\right) \equiv \overline{R}_{\mu \nu } \\
Z\left( r\right) ,\overline{Z}\left( r\right) \in C^{k}\left( \mathbb{D}%
^{4}\right)  \\
Z\left( r\right) |_{\partial \mathbb{D}^{4}}=Z_{\mathbb{D}}\left( r\right)
\\
\delta \overline{Z}_{2}\left( r\right) =0 \\
\delta \left( d\Omega \right) =d^{4}x\delta \left( \sqrt{-g}\right) \neq 0%
\end{array}%
\right\} ,  \label{Zeta_c}
\end{equation}%
where\ again $k\geq 3$ and the variational fields $\left\{ g_{\mu \nu
},R_{\mu \nu },\lambda ^{\mu \nu }\right\} $\textbf{\ }are all assumed
symmetric in the indices\textbf{\ }$\mu ,\nu $. This means, in other words,
that the Einstein-Hilbert variational principle can be replaced by an
equivalent one in which the boundary condition $\left. w^{\mu }\left( Z_{1},%
\mathcal{D}Z_{1}\right) \right\vert _{\partial \mathbb{D}^{4}}=0$ needs not
be imposed anymore, as this instead is replaced by the constraint equation (%
\ref{CONS}).

On the basis of these premises, one can prove that THM.1 reported in
Appendix holds. The notable consequence is to reproduce exactly the
Euler-Lagrange equation (\ref{EL-1}) by means of a constrained Lagrangian
variational principle. In this regard, the following comments are made:

1)\ In analogy to the customary Einstein-Hilbert principle, for both
functionals $S_{c}\left( Z,\overline{Z}\right) $ and $S_{E-c}\left( Z,%
\overline{Z}\right) $ introduced in THM.1 the continuum Lagrangian
coordinates are tensorial in character.

2)\ The asynchronous constrained principle established in THM.1 can be
extended to the case of non-vacuum Einstein equations.

3) The proposition T1$_{1}$ of the theorem can be generalized also to the
case in which the Ricci tensor $R_{\mu \nu }$ is not considered as an
independent continuum Lagrangian coordinate, but a function of the same
variational metric tensor. In fact, thanks to the presence of the constraint
and the Lagrange multiplier, any contribution arising from the Ricci tensor
vanishes in the extremal equation. This feature permits one to reach an
equivalent representation of the action in terms of $S_{E-c}\left( Z,%
\overline{Z}\right) $ (see proposition T1$_{2}$).

4) THM.1 can in principle be extended also to the Palatini approach. The
fundamental reason for this conclusion is that the second Euler-Lagrange
equation holding in such a case [i.e., Eq.(\ref{non-cov-metric})] is
identically fulfilled by the extremal continuum Lagrangian coordinates.
Hence, the variational principle (\ref{Principle}) must hold also when the
Palatini functional class (\ref{Palatini}), subject to the constraint (\ref%
{CONS}), is considered.

These conclusions show that the constrained principle given by THM.1
encompasses both the Einstein-Hilbert and Palatini approaches and provides a
convenient framework for the variational treatment of the contributions
associated with the Christoffel symbols.

\section{Gauge invariance properties}

A further critical issue inherent in certain literature formulations lies in
the lack of gauge invariance properties. This feature is related to the
adoption of a non-tensorial variational Lagrangian density $\mathcal{L}$.
This happens in particular in the case of the two variational approaches
indicated above. Indeed, this feature gives rise to\ continuum field
theories which are intrinsically non-gauge invariant. It is important to
stress, however, that the property of gauge invariance should be regarded as
a mandatory feature of variational field theories in general (see also
related discussion in Section VI ). This demands that gauge invariance
should be fulfilled both by variational ($Z$) and extremal ($\overline{Z}$\
) continuum fields, the latter being identified with the solutions of the
Euler-Lagrange equations determined by the variational principle. As a
consequence, also the variational functional $S\left( Z\right) $ and the
corresponding variational Lagrangian\textbf{\ }$L\left( Z,\mathcal{D}%
Z\right) $, together with the corresponding extremal quantities\textbf{\ }$%
S\left( \overline{Z}\right) $\textbf{\ }and\textbf{\ }$L\left( \overline{Z},%
\mathcal{D}\overline{Z}\right) $, should be necessarily determined up to a
suitable gauge contribution. However, this property is violated both in the
Einstein-Hilbert and Palatini approaches as well as in THM.1.

To illustrate the issue, consider for example the trivial gauge
transformation acting on the variational field Lagrangians considered in
these approaches:%
\begin{equation}
L\left( Z,\mathcal{D}Z\right) \rightarrow L\left( Z,\mathcal{D}Z\right) +C,
\label{wizz}
\end{equation}%
with $C$ being an arbitrary constant $4-$scalar. It follows that the
Lagrangian density $\mathcal{L}$ transforms as%
\begin{equation}
\mathcal{L}\left( Z,\mathcal{D}Z\right) \rightarrow \mathcal{L}\left( Z,%
\mathcal{D}Z\right) +\sqrt{-g}C.
\end{equation}%
It is immediate to show that, in all approaches discussed above, the
introduction of the additive constant $C$ changes in a non-trivial way the
form of the Einstein equations, which becomes%
\begin{equation}
R_{\mu \nu }-\frac{1}{2}Rg_{\mu \nu }-\frac{1}{2}Cg_{\mu \nu }=\frac{8\pi G}{%
c^{4}}T_{\mu \nu },
\end{equation}%
where $C$ plays a role analogous to the so-called cosmological constant $%
\Lambda $.

Additional difficulties may arise when a general gauge transformation of the
type%
\begin{equation}
L\left( Z,\mathcal{D}Z\right) \rightarrow L\left( Z,\mathcal{D}Z\right)
+\nabla _{\alpha }C^{\alpha }\left( Z\right)  \label{catu}
\end{equation}%
is introduced. First we notice that in the Einstein-Hilbert approach the
term $\nabla _{\alpha }C^{\alpha }\left( Z\right) $ remains truly a gauge
function, where $\nabla _{\alpha }$ must be intended as a function of the
prescribed Christoffel symbols $\Gamma _{\left( C\right) \alpha \beta }^{\mu
}\left( g\right) $. This happens because in the functional setting $\left\{
Z\right\} _{E}$ the following identity%
\begin{equation}
\nabla _{\alpha }C^{\alpha }=\frac{1}{\sqrt{-g}}\frac{\partial }{\partial
x^{\alpha }}\left( \sqrt{-g}C^{\alpha }\right)  \label{ero}
\end{equation}%
holds \cite{LL}. Hence, thanks to Gauss theorem, the same gauge term does
not contribute to the Einstein-Hilbert action functional. Instead, when the
Palatini approach is considered, the gauge function $\nabla _{\alpha
}C^{\alpha }\left( Z\right) $ must be excluded \textquotedblleft a
priori\textquotedblright , since it depends intrinsically on the variational
connections entering through the definition of $\nabla _{\alpha }$, which by
construction are considered independent of the prescribed Christoffel
symbols $\Gamma _{\left( C\right) \alpha \beta }^{\mu }\left( g\right) $.
Therefore, Eq.(\ref{ero}) does not hold anymore, so that Gauss theorem
cannot apply because the volume element of integration $d^{4}x\sqrt{-g}$ and
the differential operator $\nabla _{\alpha }$ remain in this case
independent. The implication is that both the Einstein-Hilbert and Palatini
approaches are unsatisfactory because they violate at least in part the
gauge invariance symmetries (\ref{wizz}) and (\ref{catu}). The same
conclusion can be reached for the constrained variational principles
considered in THM.1.

However, the violation of gauge invariance displayed here is a serious
inconsistency. Indeed, it is in conflict, for example, with the
gauge-invariance property of the Maxwell equations when they are considered
in the flat space-time. In such a case in fact, one finds that the
corresponding variational and field Lagrangian densities coincide, namely%
\begin{equation}
\mathcal{L}_{EM}\left( Z,\mathcal{D}Z\right) \equiv L_{EM}\left( Z,\mathcal{D%
}Z\right) .  \label{property1}
\end{equation}%
Such a property is also satisfied in curved space-time when the metric
tensor is considered extremal. Remarkably, this definition warrants that the
variational Lagrangian for the EM field has the same tensorial character
both in curved and flat space-times, so that in both cases it is actually a $%
4-$scalar. Similar considerations apply in principle also to other possible
field Lagrangians, such as those describing classical scalar fields. This
feature assures that, when the metric tensor and the connections are
extremal, the variational Lagrangian density $\mathcal{L}_{EM}\left( Z,%
\mathcal{D}Z\right) $ recovers the customary flat-space-time gauge symmetry,
so that both the transformations%
\begin{eqnarray}
\mathcal{L}_{EM} &\rightarrow &\mathcal{L}_{EM}+C,  \label{CF1} \\
\mathcal{L}_{EM} &\rightarrow &\mathcal{L}_{EM}+\nabla _{\alpha }C^{\alpha
}\left( Z\right) ,  \label{CF2}
\end{eqnarray}%
leave invariant the Maxwell equations. Notice that here the gauge $\nabla
_{\alpha }C^{\alpha }\left( Z\right) $ must be regarded as extremal both
with respect to the metric tensor and the connections.

In the following, when $\mathcal{L}\left( Z,\mathcal{D}Z\right) $ satisfies
the property $\mathcal{L}\left( Z,\mathcal{D}Z\right) =L\left( Z,\mathcal{D}%
Z\right) $, the corresponding variational principle will be referred to as
\emph{standard Lagrangian variational approach}, with $\mathcal{L}\left( Z,%
\mathcal{D}Z\right) $ being denoted as \emph{standard variational Lagrangian
density}. In such a case the following characteristic properties are
expected to be fulfilled:

1)\ The continuum field theory is manifestly covariant at all levels. In
particular, this means that the continuum Lagrangian coordinates have all a
well-defined tensorial character, so that the corresponding symbolic
Euler-Lagrange equations are manifestly covariant too.

2) The property of gauge invariance, in the sense of Eqs.(\ref{CF1})-(\ref%
{CF2}), both for the functionals and the related variational Lagrangian
densities, is warranted.

From the previous analysis a number of serious discrepancies emerges for the
variational formulations of GR considered so far and the field theories for
other classical fields, which concern the fundamental properties of manifest
covariance at all levels and gauge symmetries. The issue is whether full
consistency with these basic principles can be ultimately reached also for
the variational formulation of the Einstein equations. The achievement would
be of basic relevance for several reasons, i.e., at least:

\textit{Requirement \#1 - }To assure a unified variational treatment valid
for all classical fields in the context of GR.

\textit{Requirement \#2 -} To warrant the gauge invariance property of the
variational functional and of the related variational Lagrangian.

\textit{Requirement \#3 -} To permit the manifest covariance of the theory
at all levels and in particular to assure that both the variational
Lagrangian and the associated symbolic Euler-Lagrange equations,
corresponding to the Einstein equations, exhibit a tensorial structure and
hence are manifestly covariant too.

\section{Lagrangian variational principles in classical mechanics}

In this section we briefly summarize the basic features of variational
formulations available in relativistic classical mechanics for
single-particle Lagrangian dynamics. This analysis is useful to introduce
the concept of synchronous and asynchronous Lagrangian variational
principles, to be extended below also to classical field theory.\textbf{\ }%
In this regard, we shall denote by $r^{\mu }\left( s\right) $ the Lagrangian
world-line trajectory of a charged point particle with rest mass $m_{o}$,
charge $q_{o}$ and proper time $s$, so that the corresponding $4-$velocity
is $u^{\mu }\left( s\right) =\frac{dr^{\mu }\left( s\right) }{ds}$, while%
\begin{equation}
ds^{2}=g_{\mu \nu }(r)dr^{\nu }(s)dr^{\mu }(s).  \label{ds2}
\end{equation}%
Here the metric tensor $g_{\mu \nu }\left( r\right) $ and the Faraday tensor
$F_{\mu \nu }=\partial _{\mu }A_{\nu }-\partial _{\nu }A_{\mu }$ of the
external EM fields are considered prescribed functions of $r$, namely
extremal in the sense indicated in Section I, while omitting for brevity in
this section the barred notation.

For convenience, we start recalling the standard notions of synchronous and
asynchronous variations in the context of single-particle dynamics in GR.
For this purpose, first we consider the customary asynchronous principle,
which can be found for example in Ref.\cite{LL}. The action functional in
this case is identified with%
\begin{equation}
S_{pA}(r)=-\int\nolimits_{s_{1}}^{s_{2}}ds\left( g_{\mu \nu
}(r\left( s\right) )\frac{dr^{\nu }\left( s\right) }{ds}+qA_{\mu
}\left( r\left( s\right) \right) \right) \frac{dr^{\mu }\left(
s\right) }{ds},
\end{equation}%
where $q\equiv \frac{q_{o}}{m_{o}c^{2}}$ is the normalized charge and $s_{1}$
and $s_{2}$ are fixed boundary values. In the functional $S_{pA}(r)$, the
function $r^{\mu }\left( s\right) $ is assumed to belong to the asynchronous
functional class:%
\begin{equation}
\left\{ r^{\mu }\right\} _{A}=\left\{
\begin{array}{c}
r^{\mu }(s)\in C^{2}\left(
\mathbb{R}
\right) \\
\delta \left( ds\right) \neq 0 \\
r^{\mu }(s_{k})=r_{k}^{\mu },\ k=1,2%
\end{array}%
\right\} ,
\end{equation}%
where in particular we require%
\begin{equation}
\delta \left( ds\right) =\delta \left( \sqrt{g_{\mu \nu }(r(s))dr^{\nu
}(s)dr^{\mu }(s)}\right) .
\end{equation}%
Here $\delta $ denotes the Frechet derivative which, when acting on the
vector function $r^{\mu }\left( s\right) $, is defined simply as the virtual
displacement%
\begin{equation}
\delta r^{\mu }(s)=r^{\mu }(s)-r_{b}^{\mu }(s),
\end{equation}%
where $r^{\mu }(s)$ and $r_{b}^{\mu }(s)$ identify two arbitrary functions
belonging to the\ functional class $\left\{ r^{\mu }\right\} _{A}$. In
particular, we notice that $r_{b}^{\mu }(s)$ can be always identified with
the extremal curve $r_{extr}^{\mu }(s)$, solution of the initial-value
problem associated with the Euler-Lagrange equations given below. Hence, the
well-known asynchronous Hamilton variational principle is recovered. This is
given by the variational equation%
\begin{equation}
\delta S_{pA}(r)\equiv \left. \frac{d}{d\alpha }\Psi (\alpha )\right\vert
_{\alpha =0}=0,
\end{equation}%
to hold for arbitrary displacements $\delta r^{\mu }(s)$. Here $\Psi (\alpha
)$ is the smooth real function $\Psi (\alpha )=S_{pA}(r+\alpha \delta r)$,
being $\alpha \in \left] -1,1\right[ $ to be considered independent of $r(s)$
and $s$. The corresponding Euler-Lagrange equation becomes%
\begin{equation}
\frac{\delta S_{pA}(r)}{\delta r^{\mu }(s)}\equiv g_{\mu \nu }\frac{D}{Ds}%
\frac{dr^{\nu }\left( s\right) }{ds}-q{F_{\mu \nu }}\frac{dr^{\nu }\left(
s\right) }{ds}=0.  \label{e-l-part}
\end{equation}

Let us now consider the corresponding synchronous variational principle,
which can be found for example in Refs.\cite{Goldstein,gravi} and has been
adopted also in Refs.\cite{EPJ1,EPJ2,EPJ4,EPJ5,EPJ6,EPJ7} for the treatment
of the non-local EM interaction characterizing extended particle dynamics in
the presence of EM radiation-reaction phenomena (see also discussion below).
However, for the illustration of the theory and consistent with the purpose
of this section, in the following we restrict to the treatment of local
interactions occurring for point-like classical particles.\textbf{\ }In this
case, the functional is expressed in terms of superabundant variables $%
r^{\mu }\left( s\right) $ and $u^{\mu }\left( s\right) $ and is identified
with%
\begin{equation}
S_{pS}(r,u)=-\int\nolimits_{s_{1}}^{s_{2}}dsL_{pS}\left( r\left( s\right) ,%
\frac{dr\left( s\right) }{ds},u\left( s\right) \right) ,
\end{equation}%
where $L_{pS}$ is the $4-$scalar Lagrangian%
\begin{equation}
L_{pS}\left( r\left( s\right) ,\frac{dr\left( s\right) }{ds},u\left(
s\right) \right) \equiv \left( u_{\mu }\left( s\right) +qA_{\mu }\left(
r\left( s\right) \right) \right) \frac{dr^{\mu }\left( s\right) }{ds}-\frac{1%
}{2}u^{\mu }\left( s\right) u_{\mu }\left( s\right) ,  \notag
\end{equation}%
which is linear in $\frac{dr\left( s\right) }{ds}$. In addition, the
functions $r^{\mu }\left( s\right) $ and $u^{\mu }\left( s\right) $ are
required to belong to the synchronous functional class defined as%
\begin{equation}
\left\{ r^{\mu },u^{\mu }\right\} _{S}=\left\{
\begin{array}{c}
r^{\mu }(s),u^{\mu }\left( s\right) \in C^{2}\left(
\mathbb{R}
\right) \\
\delta \left( ds\right) =0 \\
r^{\mu }(s_{k})=r_{k}^{\mu },\ k=1,2 \\
u^{\mu }(s_{k})=u_{k}^{\mu },\ k=1,2%
\end{array}%
\right\} .
\end{equation}

Notice that here the generic functions $u^{\mu }\left( s\right) $ in $%
\left\{ r^{\mu },u^{\mu }\right\} _{S}$ are not required to satisfy the
kinematic constraint $u^{\mu }\left( s\right) u_{\mu }\left( s\right) =1$,
while the line element $ds$ is by construction required to be determined by
Eq.(\ref{ds2}) in which $r^{\mu }\left( s\right) $ is an extremal curve (see
definition below). Furthermore, here $\delta $ denotes again the variation
operator which, when acting on the functions $r^{\mu }\left( s\right) $ and $%
u^{\mu }\left( s\right) $, determines the position and velocity virtual
displacements%
\begin{eqnarray}
\delta r^{\mu }(s) &=&r^{\mu }(s)-r_{extr}^{\mu }(s),  \label{44} \\
\delta u^{\mu }(s) &=&u^{\mu }(s)-u_{extr}^{\mu }(s),  \label{44-bis}
\end{eqnarray}%
where $u_{extr}^{\mu }(s)\equiv \frac{d}{ds}r_{extr}^{\mu }(s)$, and $%
r_{extr}^{\mu }(s)$ denotes the extremal curve. Here, $r^{\mu }(s)$ and $%
u^{\mu }(s)$ are considered independent, so that $\delta r^{\mu }(s)$ and $%
\delta u^{\mu }(s)$ are independent too. In this case it is immediate to
show that the corresponding synchronous Hamilton variational principle takes
the form%
\begin{equation}
\delta S_{pS}(r,u)\equiv \left. \frac{d}{d\alpha }\Psi (\alpha )\right\vert
_{\alpha =0}=0,  \label{delta-SS}
\end{equation}%
to hold for arbitrary independent displacements $\delta r^{\mu }(s)$ and $%
\delta u^{\mu }(s)$. Here $\Psi (\alpha )$ is the smooth real function $\Psi
(\alpha )=S_{pS}(r+\alpha \delta r,u+\alpha \delta u)$, being $\alpha \in %
\left] -1,1\right[ $ to be considered independent of $r(s)$, $u\left(
s\right) $ and $s$. In this case the corresponding Euler-Lagrange equations
deliver%
\begin{eqnarray}
\frac{\delta S_{pS}(r,u)}{\delta r^{\mu }(s)} &\equiv &\frac{D}{Ds}u_{\mu }-q%
{F_{\mu \nu }}u^{\nu }=0,  \label{ss1} \\
\frac{\delta S_{pS}(r,u)}{\delta u^{\mu }(s)} &\equiv &u_{\mu }-g_{\mu \nu }%
\frac{dr^{\nu }\left( s\right) }{ds}=0,  \label{ss2}
\end{eqnarray}%
which can be combined to recover Eq.(\ref{e-l-part}) and imply also the
kinematic constraint (\textit{mass-shell constraint}) $u^{\mu }\left(
s\right) u_{\mu }\left( s\right) =1$. Notice that Eqs.(\ref{ss1}) and (\ref%
{ss2}) determine the extremal curves $r^{\mu }(s)$ and $u^{\mu }(s)$ which
belong to the functional class $\left\{ r^{\mu },u^{\mu }\right\} _{S}$ and
are solutions of the same equations.

Introducing for the particle state the symbolic representation $\mathbf{x}%
\equiv \left\{ r^{\mu },u^{\mu }\right\} $, it is possible to cast Eqs.(\ref%
{ss1}) and (\ref{ss2}) in the equivalent Lagrangian form as%
\begin{equation}
\frac{d}{ds}\mathbf{x}\left( s\right) =\mathbf{X}\left( \mathbf{x}\left(
s\right) \right) ,
\end{equation}%
where $\mathbf{X}\left( \mathbf{x}\left( s\right) \right) $ is the vector
field%
\begin{equation}
\mathbf{X}\left( \mathbf{x}\left( s\right) \right) \equiv \left\{ u^{\mu
},G^{\mu }\right\} .
\end{equation}%
In the case of EM interactions considered above, the $4-$vector $G^{\mu }$
is given by $G^{\mu }\equiv q{F_{\nu }^{\mu }}u^{\nu }$, so that $\mathbf{X}%
\left( \mathbf{x}\left( s\right) \right) $ is conservative, namely%
\begin{equation}
\frac{\partial }{\partial \mathbf{x}}\cdot \mathbf{X}\left( \mathbf{x}%
\right) \equiv \frac{\partial }{\partial u^{\mu }}G^{\mu }=0.
\label{conservative-force}
\end{equation}

The following remarks are in order regarding the comparison between the two
variational approaches given in this section:

1) A basic feature of the synchronous approach lies in the adoption of
superabundant variables from the start.

2)\ The two approaches differ for the treatment of the line element $ds$,
which is assumed to be held fixed in the synchronous principle, in the sense
that $\delta ds=0$.

3) An alternative possible definition for the synchronous functional $S_{pS}$
can be achieved based on a constrained variational principle, in terms of a
Lagrange-multiplier approach which warrants that the kinematic constraint
for $u^{\mu }\left( s\right) $ is satisfied by the extremal curves only.

4)\ The\ mass-shell constraint acting on the set of superabundant variables
is satisfied identically only by the extremal curves and not by generic
curves of $\left\{ r^{\mu },u^{\mu }\right\} _{S}$. Therefore, Eq.(\ref%
{delta-SS}) should be regarded in a proper sense as an unconstrained
variational principle.

5)\ Both for the synchronous and asynchronous functionals considered above,
the Lagrangians are $4-$scalars which are defined up to an arbitrary but
suitable gauge function, namely an exact differential. Furthermore, the
corresponding Euler-Lagrange equations are manifestly covariant. Therefore,
in both cases a standard variational Lagrangian formulation exists, which
satisfies both the principle of manifest covariance at all levels and gauge
symmetry.

6) A further notable property of the synchronous variational principle (\ref%
{delta-SS}) to be mentioned concerns the possibility of introducing
arbitrary extended phase-space transformations of the particle state $%
\mathbf{x}\equiv \left\{ r^{\mu },u^{\mu }\right\} $. This is realized by a
diffeomorphism of the form%
\begin{eqnarray}
\mathbf{x}\left( s\right) &\rightarrow &\mathbf{z}\left( s\right) \equiv
\mathbf{z}\left( \mathbf{x}\left( s\right) \right) ,  \label{1dif} \\
\mathbf{z}\left( s\right) &\rightarrow &\mathbf{x}\left( s\right) \equiv
\mathbf{x}\left( \mathbf{z}\left( s\right) \right) .  \label{2dif}
\end{eqnarray}%
Notice that, when a synchronous variation is performed on a generic curve $%
\mathbf{x}\left( s\right) $, a synchronous variation is generated on $%
\mathbf{z}\left( s\right) $, and viceversa. Then it follows that,\ denoting $%
S_{pS}(r,u)\equiv S_{pS}(x)$ and introducing the transformed functional%
\begin{equation}
S_{pS}(x\left( z\right) )=\widehat{S}_{pS}(z),
\end{equation}%
the two synchronous variational principles $\delta S_{pS}(x)=0$ and $\delta
\widehat{S}_{pS}(z)=0$, to hold respectively for arbitrary variations $%
\delta \mathbf{x}\left( s\right) $ and $\delta \mathbf{z}\left( s\right) $,
are manifestly equivalent. The variables $\mathbf{z}\left( s\right) $ are
denoted as hybrid variables, because they can differ from customary
Lagrangian coordinates and related conjugate momenta. As a consequence, it
is always possible to cast the synchronous Hamilton variational principle in
hybrid form. This feature is of basic importance, for example, for the
construction of gyrokinetic theory \cite{Bek1,Bek,Cr2013,sub1,sub2}.

A point worth to be further discussed here concerns the case of non-local
interactions investigated in Refs.\cite{EPJ1,EPJ2,EPJ4,EPJ5,EPJ6,EPJ7}.
These include EM interactions arising in $N-$body systems of point-like
relativistic particles as well as the so-called EM radiation-reaction
self-force in extended particles. In all such cases, the adoption of the
synchronous Hamilton variational principle is mandatory, since it is the
only one that permits to recover standard Lagrangian and Hamiltonian
formulation for relativistic particle dynamics (regarding the precise
definition of these notions we refer in particular to Ref.\cite{EPJ2}).

Finally, it is worth pointing out that a strong physical motivation behind
the adoption of synchronous\ rather than asynchronous variational principles
actually exists in this context. In fact, as pointed out for example in Ref.%
\cite{Goldstein}, only in the first case a Hamiltonian variational
formulation is achievable.

\section{Synchronous Lagrangian variational principles for the Einstein
equations in vacuum}

Let us now pose the problem of the construction of a Lagrangian variational
principle, which holds for the Einstein equations in vacuum and satisfies
the Requirements \#1-\#3 indicated above. We intend to show that the goal is
realized by imposing, in analogy with the discussion given on relativistic
particle dynamics, that the invariant $4-$volume element $d\Omega \equiv
d^{4}r\sqrt{-g}$ is kept fixed (i.e., prescribed) during arbitrary
variations performed on the action functional, with $d^{4}r$\ identifying
the corresponding configuration-space volume element.\textbf{\ }As a
consequence, in contrast to the variational approaches considered in Section
III, in this case $d\Omega $\ must remain constant when variations of the
fields (i.e., generalized coordinates) are performed. This feature departs
from standard approaches used in previous literature for continuum field
theories, where instead the factor $\sqrt{-g}$ is considered variational and
hence the invariant volume element is not preserved by the functional
variations. In continuum field theory, the two features are formally
analogous to particle dynamics (see discussion in Section V) as far as the
treatment of the line element $ds$ is concerned. For this reason, the two
possible routes in which $d\Omega $ is preserved or not during variations
(in the sense indicated above)\ are referred to here respectively as \emph{%
synchronous and asynchronous Lagrangian variational principles.}

In order to determine a possible realization of the new synchronous
approach, as a starting point we consider here a modified form of the
constrained Lagrangian variational principle obtained in THM.1. As a
preliminary step, invoking proposition T1$_{2}$ of the same theorem, we now
identify the functional class $\left\{ Z\right\} $ for the continuum
Lagrangian coordinates with the synchronous functional class%
\begin{equation}
\left\{ Z\right\} _{E-S}\equiv \left\{
\begin{array}{c}
Z_{1}\left( r\right) \equiv g_{\mu \nu }\left( r\right) \\
\widehat{Z}_{1}\left( r\right) \equiv \widehat{g}_{\mu \nu }\left( r\right)
\\
\widehat{Z}_{2}\left( r\right) \equiv \widehat{R}_{\mu \nu }\left( r\right)
\\
Z\left( r\right) ,\widehat{Z}\left( r\right) \in C^{k}\left( \mathbb{D}%
^{4}\right) \\
g_{\mu \nu }\left( r\right) |_{\partial \mathbb{D}^{4}}=g_{\mu \nu \mathbb{D}%
}\left( r\right) \\
\delta \widehat{Z}\left( r\right) =0 \\
\delta \left( d\Omega \right) =0 \\
g_{\mu \nu }=g^{\alpha \beta }\widehat{g}_{\alpha \mu }\widehat{g}_{\beta
\nu } \\
g^{\alpha \beta }=\widehat{g}^{\alpha \mu }\widehat{g}^{\beta \nu }g_{\mu
\nu } \\
\nabla _{\alpha }\equiv \widehat{\nabla }_{\alpha }%
\end{array}%
\right\} .  \label{Z-ES}
\end{equation}%
Here the notation is as follows. First, $Z_{1}\left( r\right) \equiv g_{\mu
\nu }\left( r\right) $ is the Lagrangian variational field. Second, $%
\widehat{Z}_{1}\left( r\right) \equiv \widehat{g}_{\mu \nu }\left( r\right) $
and $\widehat{Z}_{2}\left( r\right) \equiv \widehat{R}_{\mu \nu }\left(
r\right) $ are respectively a prescribed metric tensor (to be defined below)
and the corresponding prescribed Ricci tensor, namely the Ricci tensor
determined in terms of the same $\widehat{g}_{\mu \nu }\left( r\right) $.
Similarly, $\widehat{\nabla }_{\alpha }$ denotes the covariant derivative in
which the connections are identified with the extremal Christoffel symbols
evaluated in terms of $\widehat{g}_{\mu \nu }\left( r\right) $. This implies
in particular that the variations $\delta \widehat{Z}$ must vanish
identically, namely%
\begin{equation}
\delta \widehat{Z}\equiv 0.  \label{constraint-eq}
\end{equation}

In detail, the following additional assumptions are made:

\begin{enumerate}
\item First, $d\Omega \equiv d^{4}r\sqrt{-\widehat{g}}$, where\ $\widehat{g}$
is the determinant of $\widehat{g}_{\mu \nu }\left( r\right) $, so that $%
\delta \left( d\Omega \right) =0$. For this reason\ $\delta $ is referred to
here as the \emph{synchronous variation operator}.

\item Second, $\widehat{g}_{\mu \nu }$ and $\widehat{g}^{\mu \nu }$ are the
covariant and contravariant components of the same 2-rank tensor. By
construction we assume that:\ a)\ $\widehat{g}_{\mu \nu }$ and $\widehat{g}%
^{\mu \nu }$ lower and raise tensorial indices, so that necessarily $%
\widehat{g}_{\mu \alpha }\widehat{g}^{\mu \beta }=\delta _{\alpha }^{\beta }$%
; b) they satisfy the differential constraints $\widehat{\nabla }_{\alpha }%
\widehat{g}_{\mu \nu }=0$ and $\widehat{\nabla }_{\alpha }\widehat{g}^{\mu
\nu }=0$; c) their synchronous variations vanish identically, namely $\delta
\widehat{g}_{\mu \nu }=0$ and $\delta \widehat{g}^{\mu \nu }=0$. In other
words, $\widehat{g}_{\mu \nu }$ and $\widehat{g}^{\mu \nu }$ are considered
prescribed, namely held fixed during arbitrary synchronous variations.

\item The operator $\delta $ acts on the variational function $g^{\mu \nu }$
of $\left\{ Z\right\} _{E-S}$ in such a way to preserve its tensorial
character, namely so that%
\begin{equation}
\delta g^{\mu \nu }\left( r\right) =g^{\mu \nu }\left( r\right) -g_{1}^{\mu
\nu }\left( r\right)  \label{new-orig}
\end{equation}%
is a 2-rank tensor. This means that the difference $g^{\mu \nu }\left(
r\right) -g_{1}^{\mu \nu }\left( r\right) $ must be considered as
infinitesimal, with $g^{\mu \nu }\left( r\right) $ and $g_{1}^{\mu \nu
}\left( r\right) $ being two arbitrary functions belonging to $\left\{
Z\right\} _{E-S}$.

\item Introducing a functional of the form%
\begin{equation}
S_{1}\left( Z,\widehat{Z}\right) =\int_{\widehat{\mathbb{D}}^{4}}d\Omega
L_{1}\left( Z,\widehat{Z}\right) ,  \label{s1-diz}
\end{equation}%
we denote as \emph{synchronous variation }$\delta S_{1}\left( Z,\widehat{Z}%
\right) $ the corresponding \emph{synchronous Frechet derivative}, namely%
\begin{equation}
\delta S_{1}\left( Z,\widehat{Z}\right) \equiv \left. \frac{d}{d\alpha }\Psi
(\alpha )\right\vert _{\alpha =0},  \label{Frechet-synchronous -derivative}
\end{equation}%
where $\Psi (\alpha )$ is the smooth real function defined as $\Psi (\alpha
)=S_{1}\left( Z+\alpha \delta Z,\widehat{Z}\right) $,\emph{\ }with $\alpha
\in \left] -1,1\right[ $ to be considered independent of both $\widehat{Z}$
and\emph{\ }$r^{\mu }$.
\end{enumerate}

Then, the following result applies.

\bigskip

\textbf{THM.2 - Synchronous variational principle (vacuum Einstein equations
)}

\emph{Let us introduce the modified Einstein-Hilbert Lagrangian density, to
be identified with }$\emph{4-}$\emph{scalar variational Lagrangian density}%
\begin{equation}
L_{1}\left( Z,\widehat{Z}\right) \equiv L_{EH-S}\left( Z,\widehat{Z}\right)
h\left( Z,\widehat{Z}\right) ,  \label{lxc}
\end{equation}%
\emph{with }$Z,$ $\widehat{Z}$ \emph{belonging to synchronous functional
class }$\left\{ Z\right\} _{E-S}$ \emph{and} \emph{the corresponding action
functional (\ref{s1-diz}) to be referred to as} \emph{modified Einstein
action. Here} \emph{the notation is as follows:}\newline
A) $L_{EH-S}\left( Z,\widehat{Z}\right) $ \emph{is} \emph{the vacuum
Einstein-Hilbert Lagrangian density expressed in the functional setting }$%
\left\{ Z\right\} _{E-S}$ \emph{and thus prescribed as}%
\begin{equation}
L_{EH-S}\left( Z,\widehat{Z}\right) \equiv -\frac{c^{3}}{16\pi G}g^{\mu \nu }%
\widehat{R}_{\mu \nu }.  \label{L-EH-S}
\end{equation}%
\newline
B) \emph{Furthermore, }$h\left( Z,\widehat{Z}\right) $ \emph{is the $4-$%
scalar correction factor}%
\begin{equation}
h\left( Z,\widehat{Z}\right) \equiv \left( 2-\frac{1}{4}g^{\alpha \beta
}g_{\alpha \beta }\right) ,  \label{h1}
\end{equation}%
\emph{where in the functional setting }$\left\{ Z\right\} _{E-S}$ \emph{by
definition }$g_{\alpha \beta }=\widehat{g}_{\alpha \mu }\widehat{g}_{\beta
\nu }g^{\mu \nu }$.\newline
\emph{Then, the following propositions hold:}

T$2_{1})$ \emph{Let us introduce the modified Einstein action }$S_{1}\left(
Z,\widehat{Z}\right) $\emph{\ defined in terms of Eqs.(\ref{s1-diz}) and (%
\ref{lxc}), and the corresponding synchronous variation }$\delta S_{1}\left(
Z,\widehat{Z}\right) $\emph{. Then, the synchronous variational principle}%
\begin{equation}
\delta S_{1}\left( Z,\overline{Z}\right) =0,  \label{delta-s2}
\end{equation}%
\emph{holding for arbitrary synchronous variations }$\delta g^{\mu \nu }(r)%
\emph{,}$ \emph{determines the manifestly covariant symbolic Euler-Lagrange
equation}%
\begin{equation}
\frac{\partial L_{1}\left( Z,\widehat{Z}\right) }{\partial g^{\mu \nu }}=0,
\end{equation}%
\emph{whose solution identifies the extremal field }$g_{\mu \nu }\left(
r\right) =\overline{g}_{\mu \nu }\left( r\right) $\emph{. The previous
equation then coincides with the vacuum Einstein equations upon requiring
that }$\widehat{g}_{\mu \nu }\left( r\right) $ \emph{coincides identically
with }$\overline{g}_{\mu \nu }\left( r\right) $\emph{, namely }$\widehat{g}%
_{\mu \nu }\left( r\right) =\overline{g}_{\mu \nu }\left( r\right) $\emph{.}

T$2_{2})$ \emph{The variational Lagrangian density }$L_{1}\left( Z,\widehat{Z%
}\right) $ \emph{is gauge symmetric, namely it is defined up to arbitrary
gauge transformations} \emph{of the type}%
\begin{eqnarray}
L_{1} &\rightarrow &L_{1}+C, \\
L_{1} &\rightarrow &L_{1}+\overline{\nabla }_{\alpha }C^{\alpha }\left(
Z\right) .  \label{2-thm2}
\end{eqnarray}

\emph{Proof - }The proof of T2$_{1}$ follows by explicit evaluation of the
synchronous Frechet derivative (see definition above) in the functional
class $\left\{ Z\right\} _{E-S}$. In detail, one has that by construction
the synchronous variation of the modified Einstein action is just:%
\begin{equation}
\delta S_{1}\left( Z,\widehat{Z}\right) =\int_{\widehat{\mathbb{D}}%
^{4}}d\Omega \delta L_{1}\left( Z,\widehat{Z}\right) ,
\end{equation}%
while%
\begin{equation}
\delta L_{1}\left( Z,\widehat{Z}\right) =h\left( Z,\widehat{Z}\right) \delta
L_{EH-S}\left( Z,\widehat{Z}\right) +L_{EH-S}\left( Z,\widehat{Z}\right)
\delta h\left( Z,\widehat{Z}\right) .
\end{equation}%
Since in the functional class $\left\{ Z\right\} _{E-S}$ the Ricci tensor is
held fixed, it follows that the only terms which contribute explicitly to
the variations of the functional $\delta S_{1}\left( Z,\widehat{Z}\right) $
can be expressed in the form:%
\begin{equation}
\delta S_{1}\left( Z,\widehat{Z}\right) =\int_{\widehat{\mathbb{D}}%
^{4}}d\Omega \delta g^{\mu \nu }\left( r\right) \left[ A+B\right] =0,
\label{alfa}
\end{equation}%
where respectively%
\begin{eqnarray}
A &\equiv &h\left( Z,\widehat{Z}\right) \frac{\partial L_{EH-S}\left( Z,%
\widehat{Z}\right) }{\partial g^{\mu \nu }}, \\
B &\equiv &L_{EH-S}\left( Z,\widehat{Z}\right) \frac{\partial h\left( Z,%
\widehat{Z}\right) }{\partial g^{\mu \nu }}.
\end{eqnarray}%
Explicit calculation gives%
\begin{eqnarray}
\frac{\partial L_{EH-S}\left( Z,\widehat{Z}\right) }{\partial g^{\mu \nu }}
&=&\widehat{R}_{\mu \nu }, \\
\frac{\partial h\left( Z,\widehat{Z}\right) }{\partial g^{\mu \nu }} &=&-%
\frac{1}{2}g_{\mu \nu }.
\end{eqnarray}%
Due to the arbitrariness of $\delta g^{\mu \nu }\left( r\right) $ then it
follows%
\begin{equation}
h\left( Z,\widehat{Z}\right) \overline{R}_{\mu \nu }-\frac{1}{2}\left(
g^{\alpha \beta }\widehat{R}_{\alpha \beta }\right) g_{\mu \nu }=0.
\label{hR}
\end{equation}%
This equation coincides with the vacuum Einstein equations provided $g_{\mu
\nu }\left( r\right) =\widehat{g}_{\mu \nu }\left( r\right) =\overline{g}%
_{\mu \nu }\left( r\right) $, since by construction\emph{\ }$h\left(
Z_{1}\equiv \overline{Z}_{1}=\overline{g}^{\alpha k}\left( r\right) \right)
=1$ in such a case.

The proof of T2$_{2}$ is straightforward in the case the gauge function is
identified with the constant $C$, because synchronous variations of a
constant always vanish identically. Second, in Eq.(\ref{2-thm2}) the
covariant derivative must be considered as prescribed in terms of the fixed
metric tensor. This condition is mandatory because in the synchronous
principle the invariant volume element depends on the same prescribed metric
tensor, which is held fixed too. This represents a consistency condition for
the synchronous approach and warrants the validity of the Gauss theorem, so
that one can write%
\begin{equation}
\widehat{\nabla }_{\alpha }C^{\alpha }\left( Z\right) =\frac{1}{\sqrt{-%
\widehat{g}}}\frac{\partial }{\partial x^{\alpha }}\left( \sqrt{-\widehat{g}}%
C^{\alpha }\left( Z\right) \right) .
\end{equation}%
Therefore, the gauge invariance property follows as a direct consequence of
the boundary conditions imposed on the generalized coordinates $\left\{
Z\right\} $.

\textbf{Q.E.D.}

\bigskip

As an immediate consequence of THM.2, we can conclude that the synchronous
variational principle consistently realizes the Requirements \#1-\#3 posed
in Section III. This recovers the correct gauge invariance properties of the
theory and overcomes at once the lack of\ gauge invariance characteristic of
all asynchronous principles displayed above. In particular, in such a
framework, one has that $\mathcal{L}_{1}\left( Z,\widehat{Z}\right)
=L_{1}\left( Z,\widehat{Z}\right) $ also for the gravitational field.
According to the notation introduced here, this implies that the
corresponding variational principle defines a standard Lagrangian
variational approach.

It is important to remark that an equivalent synchronous variational
principle can in principle be obtained also adopting different choices both
for the functional class $\left\{ Z\right\} _{E-S}$ as well as the
variational action functional. A possible example is obtained by extending
the functional class $\left\{ Z\right\} _{E-S}$ in such a way that $%
g^{\alpha \beta }\left( r\right) $ and $g_{\alpha \beta }\left( r\right) $
are treated as independent superabundant Lagrangian coordinates, so that it
is not necessary to raise and lower indices in the Lagrangian density. This
is obtained by identifying the functional class with%
\begin{equation}
\left\{ Z\right\} _{E-S}\equiv \left\{
\begin{array}{c}
\left( Z_{1},Z_{2}\right) \equiv \left( g_{\mu \nu }\left( r\right) ,g^{\mu
\nu }\left( r\right) \right)  \\
\left( \widehat{Z}_{1},\widehat{Z}_{2}\right) \equiv \left( \widehat{g}_{\mu
\nu }\left( r\right) ,\widehat{g}^{\mu \nu }\left( r\right) \right)  \\
\left( \widehat{Z}_{3},\widehat{Z}_{4}\right) \equiv \left( \widehat{R}_{\mu
\nu }\left( r\right) ,\widehat{R}^{\mu \nu }\left( r\right) \right)  \\
Z\left( r\right) ,\widehat{Z}\left( r\right) \in C^{k}\left( \mathbb{D}%
^{4}\right)  \\
g_{\mu \nu }\left( r\right) |_{\partial \mathbb{D}^{4}}=g_{\mu \nu \mathbb{D}%
}\left( r\right)  \\
g^{\mu \nu }\left( r\right) |_{\partial \mathbb{D}^{4}}=g_{\mathbb{D}}^{\mu
\nu }\left( r\right)  \\
\delta \widehat{Z}\left( r\right) =0 \\
\delta \left( d\Omega \right) =0 \\
\nabla _{\alpha }\equiv \widehat{\nabla }_{\alpha }%
\end{array}%
\right\} ,
\end{equation}%
and requiring again that the fields\textbf{\ }$\left( Z\left( r\right) ,%
\widehat{Z}\left( r\right) \right) $\textbf{\ }are symmetric in the indices%
\textbf{\ }$\mu ,\nu $. Similarly, also the variational Lagrangian density
must be modified by replacing it with the symmetrized form%
\begin{equation}
L_{1-sym}\left( Z,\widehat{Z}\right) \equiv \left( -\frac{c^{3}}{16\pi G}%
\right) \frac{1}{2}\left( \widehat{R}_{\mu \nu }g^{\mu \nu }+\widehat{R}%
^{\mu \nu }g_{\mu \nu }\right) h\left( Z\right) ,
\end{equation}%
where now $h\left( Z\right) $\ depends only on variational quantities. Under
the same assumptions holding for THM.2, the resulting Euler-Lagrange
equations coincide with the mutually-equivalent covariant and contravariant
representations of the vacuum Einstein equations.

In view of THM.2, it is worth pointing out how the synchronous variational
principles determined here can be extended to take into account the presence
of a non-vanishing cosmological constant $\Lambda $. In such a case the
extremal vacuum Einstein equations become%
\begin{equation}
\overline{R}_{\mu \nu }-\frac{1}{2}\overline{R}\overline{g}_{\mu \nu
}+\Lambda \overline{g}_{\mu \nu }=0.
\end{equation}%
The synchronous variational Lagrangian density to be adopted becomes
correspondingly%
\begin{equation}
L_{1\Lambda }\equiv L_{1}\left( Z,\widehat{Z}\right) -2\Lambda h\left( Z,%
\widehat{Z}\right) .
\end{equation}%
In a similar way it is possible in principle to adopt a synchronous form of
the variational principle to treat also so-called metric $f\left( R\right) $%
-models considered in modified theories of GR \cite{fr1}.

To conclude this section, a discussion concerning the physical meaning of
the synchronous variational principle is proposed. First, it must be
remarked that the latter is actually based on the introduction of
superabundant variables. More precisely, this is realized by allowing the
prescribed metric tensor $\widehat{g}^{\beta \gamma }$\ to be independent
both of the variational and extremal metric tensors $g^{\beta \gamma }$ and $%
\overline{g}^{\beta \gamma }$, the latter being determined as a solution of
the Einstein equations. The physical interpretation of such a representation
is as follows: 1) $g_{\mu \nu }$\ is a \emph{physical continuum field},
whose dynamical equations are determined by the Euler-Lagrange equations
following from the synchronous variational principle. In this sense, $g_{\mu
\nu }$\ has no geometrical interpretation, namely it does not raise or lower
indices nor it appears in the prescribed $4-$volume element (for the same
consistency of the synchronous principle) or in the prescribed Ricci tensor $%
\widehat{R}_{\mu \nu }$.\textbf{\ }2)\ Instead, $\widehat{g}_{\mu \nu }$\
plays the role of a \emph{geometrical continuum field} in the variational
functional. It determines a number of geometric properties: the invariant $4-
$volume element, the covariant derivatives, the Ricci tensor $\widehat{R}%
_{\mu \nu }$, and finally it raises/lowers tensor indices. On the other
hand, the physical and geometrical characters which distinguish the two
fields are reconciled when the extremal condition $g_{\mu \nu }\left(
r\right) =\widehat{g}_{\mu \nu }\left( r\right) =\overline{g}_{\mu \nu
}\left( r\right) $ is set in the Euler-Lagrange equations.

\section{Synchronous variational principles for the non-vacuum Einstein
equations}

THM.2 provides a unique recipe for constructing also the appropriate form of
the field Lagrangian $L_{F}$ which carries the non-vacuum source in the
Einstein equations. The issue here is to determine however the correct
representation for $L_{F}$ which is consistent with the synchronous
variational principle, preserves the correct definition of the corresponding
stress-energy tensor and at the same time does not modify the form of the
external source field equations. Therefore, the procedure should provide in
principle, besides the non-vacuum Einstein equations, also a joint
synchronous variational principle for classical matter and the EM field.

In the literature, the Lagrangian density of the EM field is identified with%
\begin{equation}
L_{EM}\left( A_{\mu },g_{\mu \nu }\right) =-\frac{1}{16\pi c}F_{\mu \nu
}F^{\mu \nu },
\end{equation}%
where $F_{\mu \nu }=\partial _{\mu }A_{\nu }-\partial _{\nu }A_{\mu }$. This
representation is consistent with the assumptions underlying the
asynchronous principle, and in particular the requirement that the
variational $g_{\mu \nu }$ lowers and raises indices. In the context of the
synchronous variational principle instead, the variational $g_{\mu \nu }$
cannot be used to lower and raise indices. Therefore, the corresponding
synchronous Lagrangian density becomes%
\begin{equation}
L_{EM-S}\left( A_{\mu },g_{\mu \nu }\right) =-\frac{1}{16\pi c}g^{\mu \alpha
}g^{\nu \beta }F_{\mu \nu }F_{\alpha \beta }.  \label{L-EM-S}
\end{equation}%
In addition we identify the field Lagrangian density $L_{F}$ with%
\begin{equation}
L_{F}=L_{EM-S}+L_{V},  \label{lllfff}
\end{equation}%
where $L_{V}$ is a generic matter Lagrangian density, which provides the
sources for both the gravitational and EM fields. Its representation will be
specified below in Section VIII in terms of a kinetic Vlasov description of
source matter.

Here we first pose the problem of the consistent treatment of both the
variational ($L_{1F}$) and field ($L_{F}$) Lagrangians in the context of the
synchronous variational formulation developed in THM.2. The following
proposition holds.

\bigskip

\textbf{THM.3 - Stress-energy tensor in the synchronous variational principle%
}

\emph{Given validity of THM.2, the variational Lagrangian density }$L_{1F}$%
\emph{, which carries the contribution of the external sources, in the
context of the synchronous variational principle (\ref{delta-s2}) is given
by the $4-$scalar}%
\begin{equation}
L_{1F}\left( Z,\widehat{Z}\right) =L_{F}\left( Z,\widehat{Z}\right) h\left(
Z,\widehat{Z}\right) ,  \label{def-lmat}
\end{equation}%
\emph{with }$L_{F}\left( Z,\widehat{Z}\right) $ \emph{being of the form
given by Eq.(\ref{lllfff}). The corresponding action integral is therefore}%
\begin{equation}
S_{1F}\left( Z,\widehat{Z}\right) =\int_{\widehat{\mathbb{D}}^{4}}d\Omega
L_{1F}\left( Z,\widehat{Z}\right) .
\end{equation}%
\emph{As a consequence, the variational derivative }$\frac{\delta
S_{1F}\left( Z,\widehat{Z}\right) }{\delta g^{\mu \nu }}$\emph{\ must give
the correct extremal stress-energy tensor entering the non-vacuum Einstein
equations (\ref{eis}), which are defined as}%
\begin{equation}
T_{\mu \nu }\left( r\right) =-2\frac{\partial L_{1F}\left( Z,\widehat{Z}%
\right) }{\partial g^{\mu \nu }}+g_{\mu \nu }L_{1F}\left( Z,\widehat{Z}%
\right) .  \label{tensor equation}
\end{equation}%
\emph{Therefore, for the consistency of the synchronous variational
principle with the same Einstein equations, it must be}%
\begin{equation}
T_{\mu \nu }\left( r\right) =-2\frac{\delta S_{1F}\left( Z,\widehat{Z}%
\right) }{\delta g^{\mu \nu }}.
\end{equation}

\emph{Proof - }Invoking the definition (\ref{def-lmat}) and evaluating the
Frechet derivative, one obtains%
\begin{eqnarray}
\delta S_{1F}\left( Z,\widehat{Z}\right) &=&\int_{\widehat{\mathbb{D}}%
^{4}}d\Omega \delta L_{1F}\left( Z,\widehat{Z}\right)  \notag \\
&=&\int_{\widehat{\mathbb{D}}^{4}}d\Omega \left[ L_{F}\left( Z,\widehat{Z}%
\right) \delta h\left( Z,\widehat{Z}\right) +h\left( Z,\widehat{Z}\right)
\delta L_{F}\left( Z,\widehat{Z}\right) \right] ,
\end{eqnarray}%
where respectively%
\begin{eqnarray}
\delta h\left( Z,\widehat{Z}\right) &=&-\frac{1}{2}g_{\mu \nu }\delta g^{\mu
\nu }, \\
\delta L_{F}\left( Z,\widehat{Z}\right) &=&\frac{\partial L_{F}\left( Z,%
\widehat{Z}\right) }{\partial g^{\mu \nu }}\delta g^{\mu \nu }.
\end{eqnarray}%
The proof then follows by elementary algebra by recalling that, after
performing the partial derivatives, the extremal value of $h\left( \overline{%
g}_{\mu \nu }\right) =1$ must be adopted.

\textbf{Q.E.D.}

\bigskip

As a consequence of THM.3, the synchronous variational principle with the
definition (\ref{def-lmat}) preserves the correct form of the stress-energy
tensor.

\section{Kinetic description of matter source}

In this section we present an application of the theory of synchronous
variational principles developed in the previous sections. The study-case
considered concerns the treatment of matter source by means of a kinetic
description and its application to the non-vacuum variational principle for
the Einstein equations. For definiteness,\ we consider here a general
configuration corresponding to a physical system represented by a
multi-species plasma or a neutral matter. In the framework of a Vlasov
description of collisionless systems, possible realizations are provided by
kinetic equilibria arising in relativistic plasmas \cite{sub1,sub2,ko1},
dark matter halos \cite{Cr2013d}, accretion-disc plasmas \cite%
{Cr2010,Cr2011,Cr2013,PRL,Cr2013b,kovar,Cr2013c,PRE-new,APJS} and
relativistic plasmas subject to EM radiation-reaction effects \cite%
{EPJ2,maha-1,maha-2}.

To start with, we notice that the ($1-$particle) kinetic distribution
function (KDF) $f(\mathbf{x},\overline{g}^{\alpha \beta }(r),\overline{A}%
_{\gamma }(r))\equiv f(\mathbf{x})$ must be considered as held fixed (i.e.,
prescribed) in the context of a variational formulation for the non-vacuum
Einstein and Maxwell equations, since a variational formulation of the
Vlasov equation is not generally achievable when both $g_{\mu \nu }$ and $%
A_{\mu }$ are non-extremal fields in the corresponding functional
dependence. Nevertheless:

1) As shown here, both the $4-$current $J^{\mu }\left( r\right) $ and the
stress-energy tensor produced by the Vlasov source $\Pi _{\mu \nu }\left(
r\right) $ can still be considered variational in a proper sense, by a
suitable identification of the corresponding Vlasov field Lagrangian $%
L_{V}=L_{V}\left( Z,\widehat{Z},f\right) $.

2)\ We intend to point out that in such a context also the KDF can be
determined via a prescribed variational principle.

In particular, let us recall first the definitions of the extremal $4-$%
current $J^{\mu }\left( r\right) $ and stress-energy tensor $\Pi _{\mu \nu
}\left( r\right) $ in terms of the kinetic distribution function $f(\mathbf{x%
})$ \cite{degroot}. The covariant components are provided by%
\begin{eqnarray}
J_{\mu }\left( r\right) &=&\sum_{species}q_{o}c\int_{\widehat{\mathbb{U}}%
^{4}}d\eta \delta \left( u^{\alpha }u_{\alpha }-1\right) \Theta \left(
u^{0}\right) u_{\mu }f(\mathbf{x}),  \label{geimu} \\
\Pi _{\mu \nu }\left( r\right) &=&\sum_{species}m_{o}c^{2}\int_{\widehat{%
\mathbb{U}}^{4}}d\eta \delta \left( u^{\alpha }u_{\alpha }-1\right) \Theta
\left( u^{0}\right) u_{\mu }u_{\nu }f(\mathbf{x}),  \label{pai}
\end{eqnarray}%
where $d\eta \equiv d^{4}u_{\alpha }/\sqrt{-\widehat{g}}$ is the invariant $%
4-$velocity volume element, with $d^{4}u_{\alpha }\equiv
\prod\limits_{i=0,3}du_{i}$ and $\widehat{\mathbb{U}}^{4}$ is the $4-$%
velocity space.

We now introduce the following definition of the Vlasov functional as%
\begin{equation}
S_{1V}\left( Z,\widehat{Z},f\right) =\int_{\widehat{\mathbb{D}}^{4}}d\Omega
L_{1V}\left( Z,\widehat{Z},f\right) ,
\end{equation}%
where $L_{1V}\left( Z,\widehat{Z},f\right) $ is the variational Vlasov
Lagrangian density, which in accordance to the prescription given above in
THM.3, is taken of the form%
\begin{equation}
L_{1V}\left( Z,\widehat{Z},f\right) =L_{V}\left( Z,\widehat{Z},f\right)
h\left( Z,\widehat{Z}\right) ,
\end{equation}%
in which the dependence in terms of the KDF $f$\ characteristic of the
kinetic Vlasov description is explicitly displayed. In particular, here the
Vlasov-source field Lagrangian density $L_{V}\left( Z,\widehat{Z},f\right) $
is defined as follows:%
\begin{equation}
L_{V}\left( Z,\widehat{Z},f\right) =-\sum_{species}\int_{\widehat{\mathbb{U}}%
^{4}}d\eta \left[ G_{\Pi }+G_{J}\right] f(\mathbf{x})\delta (u^{j}u^{k}%
\widehat{g}_{jk}(r)-1),  \label{LLVV}
\end{equation}%
where the two $4-$scalars are%
\begin{eqnarray}
G_{\Pi } &\equiv &\frac{m_{o}c^{2}}{4}\left[ 2u_{\mu }u_{\nu }g^{\mu \nu
}(r)-h\left( Z,\widehat{Z}\right) \right] , \\
G_{J} &\equiv &\frac{1}{c^{2}}q_{o}cg^{\mu \nu }(r)u_{\nu }\left( r\right)
\left( A_{\mu }-\overline{A}_{\mu }\right) ,
\end{eqnarray}%
and the last one is defined in terms of the variational and extremal fields $%
A_{\mu }$ and $\overline{A}_{\mu }$ as indicated, to warrant that $G_{J}$ is
observable. We intend to show that these two variational contributions
provide the correct source terms for the Einstein and Maxwell equations.

In view of THMs.2-3 and the definitions for the source Lagrangian $L_{V}$,
it is possible to formulate a single synchronous variational principle
holding for classical fields, which determines both the non-vacuum Einstein
equations as well as the dynamical equation for the EM $4-$potential $A_{\mu
}$ and corresponding to the non-homogeneous Maxwell equation. This is given
by the following proposition, which extends the conclusions of THM.2.

\bigskip

\textbf{THM.4 - Non-vacuum Einstein-Maxwell synchronous variational principle%
}

\emph{Given validity of THMs.2 and 3, the total variational Lagrangian
density }$L_{tot}$\emph{, which carries the contributions of the
gravitational and EM fields as well as the additional matter and current
sources is given by the $4-$scalar}%
\begin{equation}
L_{tot}\left( Z,\widehat{Z},f\right) =L_{1}\left( Z,\widehat{Z}\right)
+L_{1F}\left( Z,\widehat{Z},f\right) ,  \label{ltot}
\end{equation}%
\emph{where }$L_{1}\left( Z,\widehat{Z}\right) $ \emph{is given by Eq.(\ref%
{lxc}), }$L_{1F}\left( Z,\widehat{Z},f\right) $\emph{\ is defined by Eq.(\ref%
{def-lmat}), with }$L_{F}\left( Z,\widehat{Z}\right) $ \emph{being
prescribed by Eq.(\ref{lllfff}) in which }$L_{EM-S}$ \emph{is defined by Eq.(%
\ref{L-EM-S}) and\ }$L_{V}\left( Z,\widehat{Z},f\right) $\emph{\ by Eq.(\ref%
{LLVV}). The Lagrangian }$L_{tot}\left( Z,\widehat{Z},f\right) $\emph{\ is
prescribed in the following synchronous functional class:}%
\begin{equation}
\left\{ Z\right\} _{tot}\equiv \left\{
\begin{array}{c}
Z_{1}\left( r\right) \equiv g_{\mu \nu }\left( r\right) \\
\left( \widehat{Z}_{1}\left( r\right) ,\widehat{Z}_{2}\left( r\right)
\right) \equiv \left( \widehat{g}_{\mu \nu }\left( r\right) ,\widehat{R}%
_{\mu \nu }\left( r\right) \right) \\
Z_{3}\equiv A_{\mu }\left( r\right) \\
Z\left( r\right) ,\widehat{Z}\left( r\right) \in C^{k}\left( \mathbb{D}%
^{4}\right) \\
Z\left( r\right) |_{\partial \mathbb{D}^{4}}=Z_{\mathbb{D}}\left( r\right)
\\
\delta \widehat{Z}\left( r\right) =0 \\
\delta \left( d\Omega \right) =0 \\
g_{\mu \nu }=g^{\alpha \beta }\widehat{g}_{\alpha \mu }\widehat{g}_{\beta
\nu } \\
g^{\alpha \beta }=\widehat{g}^{\alpha \mu }\widehat{g}^{\beta \nu }g_{\mu
\nu } \\
\nabla _{\alpha }\equiv \widehat{\nabla }_{\alpha }%
\end{array}%
\right\} ,
\end{equation}%
\emph{where }$A_{\mu }\left( r\right) $\emph{\ is subject to the Coulomb
gauge }$\nabla _{\mu }A^{\mu }=0$\emph{.\ The corresponding action integral
defined in }$\left\{ Z\right\} _{tot}$\emph{\ is therefore}%
\begin{equation}
S_{tot}\left( Z,\widehat{Z},f\right) =\int_{\widehat{\mathbb{D}}^{4}}d\Omega
L_{tot}\left( Z,\widehat{Z},f\right) .
\end{equation}%
\emph{Then, the following propositions hold:}

T$4_{1})$ \emph{The synchronous variational principle is provided by the
Frechet derivative of }$S_{tot}\left( Z,\widehat{Z},f\right) $\emph{,} \emph{%
namely }$\delta S_{tot}\left( Z,\widehat{Z},f\right) =0$\emph{,} \emph{to
hold for arbitrary independent synchronous variations of the fields }$\delta
Z$\emph{. In particular, these include:}

\emph{1)\ The variations }$\delta g^{\mu \nu }$\emph{, which provide the
non-vacuum Einstein equations (\ref{eis}) (see THMs.2 and 3).}

\emph{2)\ The variations }$\delta A_{\mu }$\emph{, which provide the
non-homogeneous Maxwell equations.}

T$4_{2})$ \emph{The total Lagrangian is defined up to an arbitrary gauge, to
be identified either with a constant real $4-$scalar }$C$ \emph{or an exact
differential of the form }$\widehat{\nabla }_{\mu }C^{\mu }\left( Z\right) $%
\emph{, with }$C^{\mu }\left( Z\right) $\emph{\ being a real $4-$vector
field.}

T$4_{3})$ \emph{The conservation law }$\overline{\nabla }_{\mu }T^{\mu \nu
}\left( r\right) =0$\emph{\ and }$\overline{\nabla }_{\mu }J^{\mu }\left(
r\right) =0$\emph{\ are identically satisfied for the extremal fields.}

\emph{Proof - }The proof of T$4_{1}$ is a consequence of THMs.2 and 3. Let
us evaluate in particular first the Einstein equations. The vacuum
contribution (i.e., the lhs of Eq.(\ref{eis})) is provided by THM.2.
Instead, the non-vacuum contribution coming from $L_{1F}$ determines the
stress-energy tensor of the general form%
\begin{equation}
T_{\mu \nu }\left( r\right) =T_{\mu \nu }^{\left( EM\right) }\left( r\right)
+T_{\mu \nu }^{\left( V\right) }\left( r\right) .
\end{equation}%
Here the first term $T_{\mu \nu }^{\left( EM\right) }\left( r\right) $
represents the customary stress-energy tensor associated with the EM field.
A straightforward calculation gives in fact%
\begin{equation}
T_{\mu \nu }^{\left( EM\right) }\left( r\right) \equiv -2\left. \frac{%
\partial \left( L_{EM-S}h\left( Z,\widehat{Z}\right) \right) }{\partial
g^{\mu \nu }}\right\vert _{extr}=\frac{1}{4\pi }\left( -F_{\mu \alpha
}F_{\nu }^{\alpha }+\frac{1}{4}F_{\alpha \beta }F^{\alpha \beta }g_{\mu \nu
}\right) ,
\end{equation}%
where on the rhs all quantities must be intended as extremal ones. Instead, $%
T_{\mu \nu }^{\left( V\right) }\left( r\right) $ arises from the variational
Vlasov Lagrangian density $L_{1V}\left( Z,\widehat{Z},f\right) $. This
gives, after explicit calculation%
\begin{equation}
T_{\mu \nu }^{\left( V\right) }\left( r\right) \equiv -2\left. \frac{%
\partial \left( L_{V}h\left( Z,\widehat{Z}\right) \right) }{\partial g^{\mu
\nu }}\right\vert _{extr}=-2\Pi _{\mu \nu }\left( r\right) ,
\end{equation}%
where $\Pi _{\mu \nu }\left( r\right) $ is defined by Eq.(\ref{pai}). In
particular, we notice that the moment associated with $G_{J}$ does not
contribute at this stage. This completes the proof for the Einstein
equations. The treatment of the Maxwell equation is analogous. In
particular, denoting $S_{EM-S}\left( Z,\widehat{Z}\right) \equiv \int_{%
\widehat{\mathbb{D}}^{4}}d\Omega L_{EM-S}h\left( Z,\widehat{Z}\right) $, the
vacuum contribution originated by $L_{EM-S}$ gives for the extremal $g_{\mu
\nu }\left( r\right) $%
\begin{equation}
\frac{\delta S_{EM-S}\left( Z,\widehat{Z}\right) }{\delta A_{\mu }\left(
r\right) }=-\frac{1}{4\pi c}\overline{\nabla }_{\nu }F^{\mu \nu }\left(
r\right) .
\end{equation}%
Instead, the Vlasov source functional gives%
\begin{equation}
\frac{\delta S_{1V}\left( Z,\widehat{Z}\right) }{\delta A_{\mu }\left(
r\right) }=-\frac{1}{c^{2}}J^{\mu }\left( r\right) ,
\end{equation}%
where $J^{\mu }\left( r\right) $ is defined by Eq.(\ref{geimu}). As a
result, the customary form of the non-homogeneous Maxwell equation is
recovered, namely%
\begin{equation}
\overline{\nabla }_{\nu }F^{\mu \nu }\left( r\right) =-\frac{4\pi }{c}J^{\mu
}\left( r\right) .  \label{maxwell}
\end{equation}%
This completes the proof of T$4_{1}$. Next, the proof of proposition T$4_{2}$
is an immediate consequence of THM.2, which extends manifestly its validity
when sources are present. Finally, concerning proposition T$4_{3}$, the
conservation laws follow from the very structure of the Einstein and Maxwell
equations.

\textbf{Q.E.D.}

\bigskip

We point out that in the present approach the form of the KDF has remained
arbitrary. This permits in principle the investigation of a variety of
physical problems ranging from kinetic equilibria to dynamically-evolving
matter distributions.

As a final target, we pose the problem of setting up a variational principle
for the Vlasov equation holding in the case of conservative vector fields $%
\mathbf{X}\left( \mathbf{x}\right) $ (see Eq.(\ref{conservative-force})). In
accordance with the assumption indicated above, we shall assume that the
Vlasov kinetic equation holds for each species belonging to the source
matter. The treatment of this issue is intrinsically related to the
variational description of single-particle dynamics presented in Section V
and permits the extension of such a theory to collisionless multi-species
systems in the framework of statistical physics.

To start with, we introduce the evolution operator (i.e., the flow $%
T_{s-s_{o}}$)\ associated with the classical dynamical system corresponding
to the $1-$particle system, which is generated by the dynamical equations (%
\ref{ss1}) and (\ref{ss2}). The initial and generic states at proper times $%
s_{o}$ and $s$ are then related by the following bijection%
\begin{equation}
\mathbf{x}\left( s_{o}\right) \equiv \mathbf{x}_{o}\leftrightarrow \mathbf{x}%
\left( s\right) ,
\end{equation}%
where in terms of the evolution operator and its inverse%
\begin{eqnarray}
\mathbf{x}\left( s\right) &=&T_{s-s_{o}}\mathbf{x}\left( s_{o}\right) ,
\label{1T} \\
\mathbf{x}\left( s_{o}\right) &=&T_{s_{o}-s}\mathbf{x}\left( s\right) .
\label{2T}
\end{eqnarray}

Then, for an arbitrary species, the Vlasov equation must hold. In the
integral (Lagrangian) form this is expressed as%
\begin{equation}
f\left( \mathbf{x}\left( s\right) \right) =f_{o}\left( \mathbf{x}\left(
s_{o}\right) \right) ,  \label{oggi}
\end{equation}%
where $f_{o}\left( \mathbf{x}\left( s_{o}\right) \right) $ is an initial KDF
at $s=s_{o}$. The implications of Eq.(\ref{oggi}) are straightforward. First
we notice that such an equation is defined only on the mass-shell. As a
consequence, for all $\mathbf{x}$ belonging to the mass-shell, there exists
a phase-space trajectory such that $\mathbf{x}\left( s\right) =\mathbf{x}$.
On the other hand, $\mathbf{x}\left( s_{o}\right) $ can be represented in
terms of $\mathbf{x}$ via Eq.(\ref{2T}). Hence, Eq.(\ref{oggi}) delivers:%
\begin{equation}
f\left( \mathbf{x}\right) =f_{o}\left( T_{s_{o}-s}\mathbf{x}\right) ,
\end{equation}%
which uniquely determines $f\left( \mathbf{x}\right) $ in terms of the
initial KDF $f_{o}$. Now, assuming without loss of generality that $f_{o}$
is a smooth differentiable function, by differentiating Eq.(\ref{oggi}) with
respect to the proper time $s$, one determines the Lagrangian differential
Vlasov equation%
\begin{equation}
\frac{d}{ds}f(\mathbf{x}\left( s\right) )=0.  \label{vlasov-lagrangian}
\end{equation}

Let us now go back to the functional $S_{1V}\left( Z,\widehat{Z},f\right) $.
Noting that again the KDF is evaluated on the mass-shell, the previous
integral representation (\ref{oggi}) can be adopted. Hence, for an arbitrary
state $\mathbf{x}$, let us consider the\textbf{\ }Lagrangian phase-space
trajectory $\mathbf{x}\left( s\right) $ having the initial condition $%
\mathbf{x}\equiv \mathbf{x}\left( s\right) $. It follows that the KDF can be
represented as%
\begin{equation}
f(\mathbf{x}\left( s\right) )=\int_{s_{o}}^{s}ds^{\prime }\left[ \frac{d}{%
ds^{\prime }}f(\mathbf{x}\left( s^{\prime }\right) )\right] +f_{o}(\mathbf{x}%
\left( s_{o}\right) ).  \label{df}
\end{equation}%
Here $\mathbf{x}\left( s_{o}\right) $ denotes the state at the proper time $%
s_{o}\neq s$ along the same trajectory. We notice that at this stage we do
not invoke yet the validity of the Vlasov equation, so that generally under
the integral $\frac{d}{ds^{\prime }}f(\mathbf{x}\left( s^{\prime }\right)
)\neq 0$, while we still must assume that single-particle dynamics can be
cast in the Lagrangian form given in Section IV. Then, let us introduce an
arbitrary smooth real phase-function $\delta s\left( \mathbf{x}\right) $. We
shall require for this purpose that $\delta s\left( \mathbf{x}\right) $
satisfies the boundary conditions $\left. \delta s\left( \mathbf{x}\right)
\right\vert _{\partial \mathbb{D}^{4}}=0$. In terms of the rhs of Eq.(\ref%
{df}), we define the variation of $\delta f(\mathbf{x})$ as%
\begin{equation}
\delta f(\mathbf{x})=\int_{s_{o}}^{s+\delta s}ds^{\prime }\left[ \frac{d}{%
ds^{\prime }}f(\mathbf{x}\left( s^{\prime }\right) )\right] ,  \label{neve}
\end{equation}%
where $\delta s\equiv \delta s\left( \mathbf{x}\right) $. Then, we perform
the variation of the functional $S_{1V}\left( Z,\widehat{Z},f\right) $ with
respect to $\delta s\left( \mathbf{x}\right) $. This must necessarily be
identified with the Frechet derivative%
\begin{equation}
\delta S_{1V}\left( Z,\widehat{Z},f\right) \equiv \left. \frac{d}{d\alpha }%
\Psi (\alpha )\right\vert _{\alpha =0},  \label{delta-vlasov-func}
\end{equation}%
where $\Psi (\alpha )$ is the smooth real function defined as $\Psi (\alpha
)=S_{1V}\left( Z,\widehat{Z},f\left( \mathbf{x}\left( s+\alpha \delta
s\right) \right) \right) $, with $\alpha \in \left] -1,1\right[ $ to be
considered independent of the variational $Z(r)$ and $\mathbf{x}$. We stress
that the definition (\ref{delta-vlasov-func}) is unique once the
identification (\ref{neve}) is made. Then, if the variational principle%
\begin{equation}
\delta S_{1V}\left( Z,\widehat{Z},f\right) =0  \label{vlasov-vp}
\end{equation}%
is required to hold for arbitrary variations $\delta s\left( \mathbf{x}%
\right) $ as well as arbitrary species, it uniquely determines the species
Vlasov equation for the corresponding (species) KDF $f(\mathbf{x})$. Notice
that in the variational principle (\ref{vlasov-vp}) above all the remaining
fields $Z$ can be considered extremal, so that in the variation the
functional $S_{1V}\left( Z,\widehat{Z},f\right) $ reduces to $S_{1V}\left( Z,%
\widehat{Z},f\right) \equiv \overline{S}_{1V}\left( \overline{Z},f\right) $,
with%
\begin{equation}
\overline{S}_{1V}\left( \overline{Z},f\right) =\int_{\widehat{\mathbb{D}}%
^{4}}d\Omega \overline{L}_{1V}\left( \overline{Z},f\right) ,
\end{equation}%
and where%
\begin{equation}
\overline{L}_{V}\left( \overline{Z},f\right) =-\frac{m_{o}c^{2}}{4}%
\sum_{species}\int_{\widehat{\mathbb{U}}^{4}}d\eta f(\mathbf{x})\delta
(u^{j}u^{k}\overline{g}_{jk}(r)-1).
\end{equation}%
Hence, the variational principle for the Vlasov equation becomes simply $%
\delta \overline{S}_{1V}\left( \overline{Z},f\right) =0$. This implies that
the variational derivative $\frac{\delta \overline{S}_{1V}\left( \overline{Z}%
,f\right) }{\delta s\left( \mathbf{x}\right) }$ recovers the correct Vlasov
equation written in Lagrangian form, namely Eq.(\ref{vlasov-lagrangian}).

To conclude the analysis, we notice that the variational principle (\ref%
{vlasov-vp}) defined above is synchronous with respect to the phase-space
volume element $d\Omega d\eta $, because the latter remains unchanged during
the variation. For conservative dynamical systems, the present variational
principle represents an alternative to the approach described in Ref.\cite%
{morrison}.

\section{Conclusions}

In this paper the problem of formulating synchronous Lagrangian variational
approaches in the context of General Relativity has been addressed. The
motivations of the investigations are related to two different critical
aspects of previous literature approaches to the variational formulations of
GR. The first one is the lack of a manifestly-covariant variational theory,
in which both the Lagrangian variables and the symbolic Euler-Lagrange
equations are tensorial in character. The second issue is related to the
violation of gauge symmetry arising in the same formulations.

In this regard, a number of results has been achieved. Starting from the
analysis of historical literature on the variational principles for the
Einstein equations, the origins of the aforementioned difficulties have been
analyzed. In particular, it has been shown that the Einstein-Hilbert and
Palatini variational approaches can be summarized in a single theorem
realized by means of as constrained variational principle. Its various
possible equivalent realizations have been discussed in THM.1.

In order to overcome these limitations, a preliminary discussion has
concerned the variational principles in relativistic classical mechanics. In
such a context two different variational approaches are available,
respectively based on asynchronous and synchronous Hamilton variational
principles. The basic difference between the two approaches lies on the
constraint to be placed on the line element $ds$, which is considered
variational in the case of asynchronous principles, and extremal (i.e.,
prescribed during variations)\ for synchronous ones. This feature suggests
in a natural way its possible extension in the context of continuum
classical fields. As shown in this paper, this is achieved by means of the
introduction of constrained synchronous variational principles, in which the
configuration-space $4-$scalar volume element $d\Omega $ is left invariant
by this kind of variations. This approach differs from those typically
adopted in the literature, in which instead the volume element is not
preserved during variations, and are therefore referred to here as
asynchronous principles.

A number of related theorems have been proved. The first one (THM.2) deals
with the vacuum Einstein equations, for which a synchronous Lagrangian
variational principle has been established. As discussed in the paper, this
involves necessarily the adoption of a constrained formulation, which
warrants the conservation of the $4-$scalar volume element. This demands
however that also the Christoffel symbols must be considered prescribed,
i.e., fixed, during synchronous variations. In turn this implies that both
the covariant derivative and the Ricci tensor itself must be considered as
prescribed in the same way.

The theorem has several remarkable implications. The first one concerns the
physical interpretation of the variational constraints. Indeed, the
synchronous principle requires the existence of a prescribed field $\widehat{%
g}_{\mu \nu }\left( r\right) $ which in a sense effectively determines the
geometric properties associated with the functional setting, namely which
prescribes the volume element as well as the Ricci curvature tensor and the
covariant derivative operator, and at the same time raises and lowers tensor
indices. Then, as shown here, the Einstein equations are recovered once the $%
\widehat{g}_{\mu \nu }\left( r\right) $ is identified with the extremal
metric tensor. A further physically-meaningful aspect of the theory concerns
the fulfillment of both gauge symmetry and the property of manifest
covariance of the theory. The analysis of the gauge invariance property of
this approach is revealing. This concerns general gauge functions which can
be either arbitrary constants or exact differentials. Its validity in fact
is a unique consequence of the adoption of the synchronous variational
principle, because in such a case the variational Lagrangian and the
Lagrangian density coincide and are realized in terms of a $4-$scalar field.
On the contrary, asynchronous principles to be found in the previous
literature are never fully gauge invariant, since the variational Lagrangian
density is not a $4-$scalar. As a side consequence, this feature prevents
the possibility of satisfying the manifest covariance also in the case of
symbolic Euler-Lagrange equations stemming from asynchronous variational
principles.

As a further interesting development the case of the non-vacuum Einstein and
Maxwell equations have been treated in terms of a single synchronous
variational principle. For this purpose, first it has been proved that such
a principle preserves the correct prescription of the stress-energy tensor
expected for the source fields (THM.3). Then, a joint synchronous
variational principle has been established both for the Einstein and Maxwell
equations. Its basic feature has been shown again to realize the properties
of manifest covariance and gauge symmetry even in the presence of classical
external sources (THM.4). In this reference, the case of a matter source
described in the framework of a Vlasov kinetic theory has been considered
and the corresponding Vlasov variational Lagrangian density uniquely
determined. As a notable consequence, it has been shown that in terms of the
action functional, a variational principle can be achieved also for the
Vlasov equation. The latter applies in the case of particle dynamics subject
to a conservative $4-$vector force.

The theory presented here exhibits intriguing features, providing a novel
route for the variational treatment of classical continuum field theory in
the context of General Relativity. In authors' view the new variational
approach might be susceptible of developments concerning the Hamiltonian
treatment of General Relativity and quantum gravity.

\section{Acknowledgments}

Work developed within the research projects of the Czech Science Foundation
GA\v{C}R grant No. 14-07753P (C.C.)\ and Albert Einstein Center for
Gravitation and Astrophysics, Czech Science Foundation No. 14-37086G (M.T.).
The authors acknowledge the hospitality of the International Center for
Theoretical Physics (ICTP, Trieste, Italy) during the final stage of
preparation of the manuscript.

\section{Appendix - Constrained variational principle for the Einstein
equations}

In this Appendix we report the proof of a theorem concerning an alternative
constrained variational principle of asynchronous type for the Einstein
equations. This is given by adopting the functional class defined by Eq.(\ref%
{Zeta_c}).

\bigskip

\textbf{THM.1 - Constrained Einstein-Hilbert variational principle}

\emph{Let us consider for simplicity and without loss of generality the case
of the vacuum Einstein equations. We define the constrained action integral }%
$S_{c}$ \emph{as}%
\begin{equation}
S_{c}\left( Z,\overline{Z}\right) =\int_{\widehat{\mathbb{D}}^{4}}d^{4}x%
\sqrt{-g}\left[ L_{c}\left( Z\right) +\lambda ^{\mu \nu }\left( R_{\mu \nu }-%
\overline{R}_{\mu \nu }\right) \right] ,
\end{equation}%
\emph{where }$L_{c}\left( Z\right) $\emph{\ is defined as}%
\begin{equation}
L_{c}\left( Z\right) \equiv -\frac{c^{3}}{16\pi G}g^{\mu \nu }R_{\mu \nu },
\end{equation}%
\emph{with }$R_{\mu \nu }$\emph{\ and }$g^{\mu \nu }$\emph{\ to be
considered as independent variational functions. In addition, }$\lambda
^{\mu \nu }$\emph{\ is a Lagrange multiplier to be included in }$\left\{
Z\right\} $\emph{\ and }$R_{\mu \nu }$\emph{,} $\overline{R}_{\mu \nu }$
\emph{are the variational and extremal Ricci tensors, the latter being
defined in particular by Eq.(\ref{rirr}) and evaluated in terms of the
extremal metric tensor }$\overline{g}_{\mu \nu }$\emph{. Here the functional
class is defined as} \emph{in Eq.(\ref{Zeta_c}).} \emph{Then, the following
propositions hold:}

T1$_{1})$\emph{\ The variational principle}%
\begin{equation}
\delta S_{c}\left( Z,\overline{Z}\right) =0,  \label{Principle}
\end{equation}%
\emph{to hold for arbitrary variations of the generalized coordinates }$%
\delta Z$\emph{\ is such that: a) it yields as extremal equations the vacuum
Einstein equations, once }$g_{\mu \nu }=\overline{g}_{\mu \nu }$ \emph{is
identified with the extremal metric tensor; b)\ it determines the extremal
value }$R_{\mu \nu }$\emph{, which must coincide with }$\overline{R}_{\mu
\nu }$\emph{;\ c)\ it yields the extremal value of }$\lambda ^{\mu \nu }$%
\emph{\ as }$\overline{\lambda }^{\mu \nu }=\frac{c^{3}}{16\pi G}\overline{g}%
^{\mu \nu }$\emph{.}

T1$_{2})$\emph{\ The action functional can be equivalently replaced with }%
\begin{equation}
S_{E-c}\left( Z,\overline{Z}\right) =\int_{\widehat{\mathbb{D}}^{4}}d^{4}x%
\sqrt{-g}L_{E-c}\left( Z,\overline{Z}\right) ,
\end{equation}%
\emph{to be denoted as constrained Einstein action, where now}%
\begin{equation}
L_{E-c}\left( Z,\overline{Z}\right) \equiv -\frac{c^{3}}{16\pi G}g^{\mu \nu }%
\overline{R}_{\mu \nu },
\end{equation}%
\emph{and the functional class becomes}%
\begin{equation}
\left\{ Z\right\} _{E-c}\equiv \left\{
\begin{array}{c}
Z\left( r\right) \equiv g_{\mu \nu }\left( r\right) \\
\overline{Z}\left( r\right) \equiv \overline{R}_{\mu \nu } \\
Z\left( r\right) ,\overline{Z}\left( r\right) \in C^{k}\left( \mathbb{D}%
^{4}\right) \\
Z\left( r\right) |_{\partial \mathbb{D}^{4}}=Z_{\mathbb{D}}\left( r\right)
\\
\delta \overline{Z}\left( r\right) =0 \\
\delta \left( d\Omega \right) =d^{4}x\delta \left( \sqrt{-g}\right) \neq 0%
\end{array}%
\right\} ,
\end{equation}%
\emph{with }$k\geq 3$\emph{, while both }$g_{\mu \nu }\left( r\right) $\emph{%
\ and }$\overline{R}_{\mu \nu }$\emph{are assumed symmetric in the indices} $%
\mu ,\nu $\emph{.}

\emph{Proof -- }Let us start from proposition T1$_{1}$. The variation of the
Lagrange multiplier gives%
\begin{equation}
\frac{\delta S_{c}\left( Z,\overline{Z}\right) }{\delta \lambda ^{\mu \nu }}%
=R_{\mu \nu }-\overline{R}_{\mu \nu }=0,  \label{tt1}
\end{equation}%
which provides the representation for the extremal Ricci tensor in terms of
the Christoffel symbols carried by $\overline{R}_{\mu \nu }$. Then, the
independent variation with respect to $R_{\mu \nu }$ yields%
\begin{equation}
\frac{\delta S_{c}\left( Z,\overline{Z}\right) }{\delta R_{\mu \nu }}=-\frac{%
c^{3}}{16\pi G}g^{\mu \nu }+\lambda ^{\mu \nu }=0,  \label{tt2}
\end{equation}%
which determines the extremal value of the Lagrange multiplier. Finally, the
independent variation with respect to $g^{\mu \nu }$ gives%
\begin{equation}
\frac{\delta S_{c}\left( Z,\overline{Z}\right) }{\delta g^{\mu \nu }}\equiv
\frac{\partial \left[ \sqrt{-g}L_{c}\left( Z\right) \right] }{\partial
g^{\mu \nu }}=0,
\end{equation}%
where we have already taken into account the validity of Eq.(\ref{tt1}).
This equation can be written explicitly as%
\begin{equation}
\frac{\partial \left[ \sqrt{-g}L_{c}\left( Z\right) \right] }{\partial
g^{\mu \nu }}=\overline{R}_{\mu \nu }-\frac{1}{2}\left( g^{\alpha \beta }%
\overline{R}_{\alpha \beta }\right) g_{\mu \nu }=0,
\end{equation}%
which recovers the vacuum Einstein equations upon identifying $g_{\mu \nu }$
with the extremal metric tensor. This completes the proof of T1$_{1}$.

As for proposition T1$_{2}$, one obviously has%
\begin{equation}
\frac{\delta S_{E-c}\left( Z,\overline{Z}\right) }{\delta g^{\mu \nu }}=%
\frac{\delta S_{c}\left( Z,\overline{Z}\right) }{\delta g^{\mu \nu }}=0,
\end{equation}%
which recovers the Einstein equations in vacuum, in which $\overline{R}_{\mu
\nu }$ is already represented in terms of the extremal metric tensor.

\textbf{Q.E.D.}

\bigskip

\bigskip

\end{document}